\documentclass[%
aps,nofootinbib,notitlepage,superscriptaddress,10Limberkipt,prd, twocolumn,
 amsmath,amssymb
]{revtex4-2}

\usepackage[caption=false]{subfig}
\usepackage{braket}
\usepackage{float}
\usepackage{lipsum}
\usepackage{graphicx}
\usepackage{dcolumn}
\usepackage{bm}
\usepackage{natbib}
\usepackage{hyperref}
\hypersetup{
	colorlinks = true,
    linkcolor = Maroon,
    urlcolor  = Maroon,
    citecolor = Maroon
}
\usepackage{afterpage}
\usepackage{placeins}
\usepackage{microtype}
\usepackage{verbatim}
\usepackage[amssymb]{SIunits}
\usepackage{tabularx}
\usepackage[dvipsnames]{xcolor}
\usepackage{tikz}

\usepackage{color}
\definecolor{darkgreen}{RGB}{0,120,0}

\newcommand{\av}[1]{\langle{#1}\rangle}

\newcommand{\be}{\begin{equation}}
\newcommand{\ee}{\end{equation}}
\newcommand{\bea}{\begin{eqnarray}}
\newcommand{\eea}{\end{eqnarray}}
\newcommand{\beas}{\begin{eqnarray*}}
\newcommand{\eeas}{\end{eqnarray*}}

\newcommand{\vx}{\mathbf{x}}
\newcommand{\vk}{\mathbf{k}}
\newcommand{\vq}{\mathbf{q}}
\newcommand{\vp}{\mathbf{p}}

\newcommand{\bx}{{\boldsymbol x}}

\def\gsim{ \lower .75ex \hbox{$\sim$} \llap{\raise .27ex \hbox{$>$}} }
\def\lsim{ \lower .75ex \hbox{$\sim$} \llap{\raise .27ex \hbox{$<$}} }

\def\dalam{\hbox
{\vrule\vbox{\hrule\hbox to 1ex{ \hfill}\kern 1 ex\hrule}\vrule}}

\def\1/2{\hbox{$ {1 \over 2}$ }}

\begin{document}

\title{{\Large Wonderings on Wiggly Bispectra:}\\Non-linear Evolution and Reconstruction of Oscillations in the Squeezed Bispectrum}

\author{Samuel~Goldstein}
\email{sjg2215@columbia.edu}
\affiliation{Department of Physics, Columbia University, New York, NY 10027, USA}

\author{Oliver~H.\,E.~Philcox}
\affiliation{Department of Physics, Columbia University, New York, NY 10027, USA}
\affiliation{Simons Society of Fellows, Simons Foundation, New York, NY 10010, USA}
\affiliation{Department of Physics,
Stanford University, Stanford, CA 94305, USA}
\author{Emanuele~Fondi}
\affiliation{Institut de Ciències del Cosmos, Universitat de Barcelona (ICCUB), Martí i Franquès, 1, 08028 Barcelona, Spain}
\author{William~R.~Coulton}
\affiliation{Kavli Institute for Cosmology Cambridge, Madingley Road, Cambridge CB3 0HA, UK}
\affiliation{DAMTP, Centre for Mathematical Sciences, University of Cambridge, Cambridge CB3 OWA, UK}

\begin{abstract}
\noindent Oscillations in the primordial bispectrum are sourced by a range of inflationary phenomena, including features in the inflaton potential and interactions with massive fields through the Cosmological Collider scenario. These signatures offer a powerful window into early-universe physics. In this work, we study how oscillations of the form $\lim_{q\ll k}B(q,k)\propto \cos(\mu \ln(q/k))$ impact the non-linear squeezed matter bispectrum. Using a suite of $N$-body simulations with non-Gaussian initial conditions, we show that non-linear evolution significantly damps these oscillations, effectively erasing the signal on scales $k \gtrsim 0.3~h/{\rm Mpc}$ at redshift $z=0$. This damping is well-described by the Zel'dovich approximation and can be modeled deep into the non-linear regime using non-perturbative separate universe simulations. Promisingly, we show that reconstruction techniques developed for baryon acoustic oscillation (BAO) analyses can largely undo this damping, improving constraints on the amplitude (phase) of oscillations in the primordial squeezed bispectrum by up to a factor of five (four) at $z=0$. We also discuss several challenges with modeling the non-linear evolution of the squeezed bispectrum in the Cosmological Collider scenario, where the bispectrum is suppressed by a factor of $(q/k)^{3/2}$ relative to the template studied here. Our findings pave the way for future searches for oscillatory bispectra using large-scale structure data. 
\end{abstract}

\maketitle

\section{Introduction}\label{Sec:Intro}

\noindent Uncovering the fundamental physics responsible for inflation is a central goal of modern cosmology. In the simplest single-field slow-roll scenarios, the primordial perturbations are Gaussian and characterized by a nearly scale-invariant power spectrum~\cite{Maldacena:2002vr, Creminelli:2004yq, Creminelli:2011rh, Pajer:2013ana}. However, more complex inflationary scenarios, such as those involving additional fields or features in the inflaton potential, can break scale invariance and/or generate non-Gaussianities in the primordial perturbations~\cite{Meerburg:2019qqi, Achucarro:2022qrl}. While current observations are consistent with Gaussian initial conditions~\cite{Planck:2019kim, Jung:2025nss}, ongoing and upcoming cosmological surveys are expected to provide unprecedented sensitivity to primordial non-Gaussianity~\cite{EUCLID:2011zbd, DESI:2013agm,SPHEREx:2014bgr, LSSTDarkEnergyScience:2018jkl, SimonsObservatory:2018koc, CMB-S4:2016ple}.

The main target for non-Gaussianity searches is the primordial bispectrum. A vast range of bispectra can be generated depending on the inflationary model~\cite{Babich:2004gb}. Here, we focus on oscillatory bispectra, which can arise from, \emph{e.g.}, sharp features in the inflaton potential~\cite{Chen:2008wn}, resonances due to periodic modulations in the inflaton potential~\cite{Chen:2008wn, Flauger:2009ab, Chen:2010bka}, and non-Bunch Davies initial conditions~\cite{Meerburg:2009ys, Meerburg:2009fi}. Additionally, couplings between the inflaton and massive fields with $m>3H/2$, where $H$ is the inflationary Hubble scale, introduce oscillations in the squeezed bispectrum with a frequency that depends on the mass and spin of the field~\cite{Arkani-Hamed:2015bza,Assassi:2012zq,Chen:2009zp,Lee:2016vti}. This  ``Cosmological Collider" scenario is a powerful probe of massive fields in the early Universe~\cite{Chen:2009zp, Chen:2009we, Baumann:2011nk, Chen:2012ge, Noumi:2012vr, Assassi:2012zq, Ghosh:2014kba, Arkani-Hamed:2015bza, Schmidt:2015xka, Dimastrogiovanni:2015pla, Gleyzes:2016tdh, Lee:2016vti, Kehagias:2017cym, An:2017hlx, Kumar:2017ecc,MoradinezhadDizgah:2017szk, Baumann:2017jvh, Bordin:2018pca, Cabass:2018roz, Kumar:2018jxz,  MoradinezhadDizgah:2018ssw, Goon:2018fyu, Hook:2019zxa, Kumar:2019ebj, Liu:2019fag, Wang:2019gbi, Akitsu:2020jvx, Bodas:2020yho, Lu:2021wxu, Pinol:2021aun, Cui:2021iie, Reece:2022soh, Jazayeri:2022kjy, Chen:2022vzh, Chen:2023txq, Werth:2023pfl, Jazayeri:2023xcj, Chakraborty:2023qbp, Chakraborty:2023eoq, Pinol:2023oux, Green:2023uyz, Adshead:2024paa, Sohn:2024xzd, Cabass:2024wob,  McCulloch:2024hiz, Quintin:2024boj, Sohn:2024xzd, Goldstein:2024bky, Philcox:2025bvj, Philcox:2025lrr, Philcox:2025wts}. In short, there are strong theoretical reasons to search for oscillatory bispectra.

Currently, the tightest constraints on a range of bispectrum templates come from the cosmic microwave background (CMB)~\cite{Planck:2019kim, Jung:2025nss}; however, near-term large-scale structure (LSS) surveys have the potential to surpass existing CMB bounds~\cite{Sailer:2021yzm, Cabass:2022epm, Spec-S5:2025uom}. To fully realize this potential, it is essential to understand the complexities that arise in LSS analyses, such as non-linear structure formation, redshift space distortions, and galaxy formation and biasing.\footnote{Even in the linear regime, LSS constraints on primordial non-Gaussianity can be sensitive to non-linearities. For example, constraints on $f_{\rm NL}^{\rm loc}$ using the scale-dependent bias require knowledge about the non-Gaussian bias $b_\phi,$ which can be a challenging game of galactic guesswork~\cite{Barreira:2020kvh, Barreira:2021ueb, Barreira:2022sey}.} This is particularly true for oscillatory signatures, which can be severely damped by large-scale bulk flows and non-linear structure formation.

While there has been extensive work focused on modeling and reconstructing oscillations in the power spectrum, both in the context of baryon acoustic oscillations (BAO) and primordial features~\cite{Meiksin:1998ra, Seo:2005ys, White:2005tf, Eisenstein:2006nk, Padmanabhan:2008dd, Sherwin:2012nh, Vlah:2015zda, Beutler:2019ojk, Ballardini:2019tuc, Chen:2020ckc, Euclid:2023shr, Chen:2024tfp, Ballardini:2024dto, Calderon:2025xod, Stahl:2025qru}, oscillatory bispectra have received considerably less attention. Refs.~\cite{Cyr-Racine:2011bjz, MoradinezhadDizgah:2017szk, MoradinezhadDizgah:2018ssw, Cabass:2018roz} derived models for the scale-dependent bias and the galaxy bispectrum for several oscillatory bispectra. More recently, Ref.~\cite{Behera:2023uat} studied BAO damping in the bispectrum using $N$-body simulations, and Ref.~\cite{Chen:2024pyp} used Lagrangian perturbation theory to study the damping of oscillatory features in the bispectrum, particularly those sourced by oscillations in the primordial power spectrum. Nevertheless, to our knowledge, there have been no attempts to run $N$-body simulations including an oscillatory primordial bispectrum. Here, we attempt precisely this, using simulations to study the non-linear evolution of wiggles in the primordial squeezed bispectrum.

Herein, we run $N$-body simulations with a bispectrum template \emph{inspired} by the Cosmological Collider scenario. This template peaks in the local shape, but is modulated by logarithmic oscillations in the ratio of $q/k,$ where $q$ and $k$ are the long-wavelength (soft) and short-wavelength (hard) modes, respectively, \emph{i.e.}, $\lim_{q\ll k}\langle \zeta(q)\zeta(k)\zeta(k')\rangle \propto \cos(\mu\ln(q/k))P_\zeta(q)P_\zeta(k)$. We find that these oscillations are severely damped by non-linear structure formation and that this damping can be effectively modeled in the quasi-linear regime using the Zel'dovich approximation. Building on recent advances in non-perturbative modeling of the squeezed bispectrum~\cite{Goldstein:2022hgr, Giri:2023mpg, Goldstein:2023brb, Goldstein:2024bky}, we develop and validate a model for this damping using separate universe simulations. Using this model, we recover unbiased constraints on both the amplitude and phase of the primordial oscillations from the non-linear squeezed matter bispectrum across a broad range of scales and redshifts. Furthermore, motivated by the success of reconstruction techniques in restoring oscillatory features in the power spectrum, we apply an iterative reconstruction algorithm to our simulations. This procedure largely restores the primordial oscillations in the bispectrum, substantially increasing the detectability of these features. Finally, we comment on several challenges in extending these results to the real Cosmological Collider squeezed bispectrum, which is suppressed by a factor of $(q/k)^{3/2}$ relative to the template used here.

The remainder of this paper is organized as follows. In Sec.~\ref{Sec:background}, we introduce the bispectrum template used in this work and derive theoretical predictions for the non-linear matter bispectrum associated with this model. In Sec.~\ref{Sec:methodology}, we discuss our methodology and analysis pipeline. In Sec.~\ref{Sec:results}, we present our results. In Sec.~\ref{Sec:conclusions}, we summarize our findings and highlight possible future avenues of exploration. Appendix~\ref{App:O_fnlsq_contribution} provides details on the $\mathcal{O}(f_{\rm NL}^2)$ contributions to the squeezed bispectrum in our $N$-body simulations, and our procedure for subtracting contributions from the coupling of $f_{\rm NL}$ and gravity to the bispectrum in Fig.~\ref{fig:full_bispectrum_meas_Delta0}. In Appendix~\ref{App:Delta_gtr_0}, we present results from separate universe simulations with the Cosmological Collider $(q/k)^{3/2}$ momentum scaling, highlighting some of the challenges in extending our analysis to this scenario.

\section{Theoretical background}\label{Sec:background}
\subsection{Background and conventions}

\noindent We use the following Fourier convention:
\begin{equation}
    \delta({\bx})= \int \frac{d^3k}{(2\pi)^3}\,\delta(\vk)e^{-i\vk \cdot \bx}\equiv \int_{\vk}\delta(\vk)e^{-i\vk \cdot \bx}.
\end{equation}
For a three-dimensional statistically homogeneous and isotropic field $\delta(\vk)$,  the power spectrum and bispectrum are defined by
\begin{align}
    \langle \delta(\vk_1)\delta(\vk_2)\rangle'_{c} &\equiv P(k_1)\,, \\
    \langle \delta(\vk_1) \delta(\vk_2) \delta(\vk_3) \rangle'_{c} &\equiv B(k_1, k_2,k_3)\,,
\end{align}
respectively, where $\langle \dots \rangle'_c$ denotes the connected correlator without the overall momentum-conserving delta function. The bispectrum is parameterized by the side lengths of three momentum vectors forming a triangle. We focus on the so-called \emph{squeezed} bispectrum, where one of the momenta, denoted by $q$, is much smaller than the other two, $k \approx k’$, and write the squeezed bispectrum as $B(\vq, \vk)$.

We work in natural units and assume a $\Lambda$CDM cosmology consistent with the fiducial parameters of the \textsc{Quijote} simulations~\cite{Villaescusa-Navarro:2019bje}: $\Omega_m = 0.3175$, $\Omega_b = 0.049$, $h = 0.6711$, $n_s = 0.9624$, and $\sigma_8 = 0.834$.

\subsection{Oscillatory bispectrum template}
\noindent In this work, we study primordial non-Gaussianity described by a bispectrum that oscillates in the squeezed limit as follows,
\begin{equation}\label{eq:squeezed_bk_osc}
B_{\Phi}(\vq,\vk) = 4 f_{\rm NL}\left(\frac{q}{k}\right)^\Delta\cos\left[\mu\ln(q/k)+\delta\right] P_{\Phi}(q)P_{\Phi}(k),
\end{equation}
where $f_{\rm NL}$ is the amplitude of non-Gaussianity,\footnote{The normalization is chosen such that Eq.~\eqref{eq:squeezed_bk_osc} reduces to the standard local shape when $\mu=\delta=0$.} $0\leq \Delta\leq 3/2$ sets the power-law scaling, $\mu$ is a frequency, $0\leq \delta<\pi$ is a phase,\footnote{We only consider phases up to $\pi$ as we allow $f_{\rm NL}$ to be negative.} and $P_{\Phi}$ is the power spectrum of the primordial Bardeen potential, $\Phi$.

Eq.~\eqref{eq:squeezed_bk_osc} is a phenomenological template that we adopt to isolate the impact of logarithmic oscillations in the squeezed bispectrum on LSS observables. It is closely related to the squeezed bispectrum produced in the Cosmological Collider scenario, whereby interactions between the inflaton and a massive scalar field (with mass $m>3H/2$) produce a bispectrum that oscillates logarithmically in $q/k$~\cite{Arkani-Hamed:2015bza,Assassi:2012zq,Chen:2009zp,Lee:2016vti}. In this case, $\Delta=3/2$, and the bispectrum is heavily suppressed. Additionally, the frequency is set by the mass, $\mu = \sqrt{m^2/H^2 - 9/4}$, and the phase depends on the precise details of the massive field and its coupling to the inflaton~\cite{Lee:2016vti,MoradinezhadDizgah:2017szk, MoradinezhadDizgah:2018ssw, Sohn:2024xzd}. The collider model can also include an angular dependence for higher-spin fields. 

Here, we focus primarily on oscillatory bispectra with $\Delta=0$, as our main goal is not to model the Cosmological Collider bispectrum, but to understand how oscillations in the squeezed bispectrum evolve under non-linear structure formation. However, where relevant, we highlight several challenges that can arise when including the steep $(q/k)^{3/2}$ scaling of the Cosmological Collider scenario, and we present a preliminary investigation of this model in Appendix~\ref{App:Delta_gtr_0}. We also note that Ref.~\cite{Goldstein:2024bky} ran $N$-body simulations with non-oscillatory squeezed bispectra in cosmologies with the $\left(q/k\right)^\Delta$ suppression for $\Delta=0.5$ and $\Delta=1.0$. Similarly, Ref.~\cite{Akitsu:2020jvx} ran $N$-body simulations with the angular dependence sourced by a massless spin-2 field.

We also note that inflationary models with sharp or periodic features in the inflaton potential can lead to oscillations in the bispectrum that are linear or logarithmic in the total momentum, respectively~\cite{Chen:2008wn, Flauger:2010ja, Chen:2010bka}. However, in accordance with the single-field consistency condition~\cite{Maldacena:2002vr, Creminelli:2004yq, Creminelli:2011rh, Pajer:2013ana},\footnote{Single-field inflationary models that deviate from the attractor limit, such as ultra-slow-roll inflation~\cite{Tsamis:2003px, Kinney:2005vj}, can violate the single-field consistency relation~\cite{Tsamis:2003px, Kinney:2005vj, Namjoo:2012aa, Martin:2012pe, Suyama:2021adn}.} the squeezed bispectrum generated in such scenarios is typically strongly suppressed, with the leading order contribution starting at $\mathcal{O}\left(\frac{q^2}{k^2} \right)$ (see, \emph{e.g.}, Refs.~\cite{Creminelli:2011rh, Cabass:2018roz}). As a result, we do not consider these models here because their signatures typically peak in non-squeezed configuration. These models can be simulated using similar techniques to the methods developed here.

\subsection{Non-linear squeezed matter bispectrum}\label{subsec:squeezed_bispectrum_deriv}
\noindent The squeezed bispectrum in Eq.~\eqref{eq:squeezed_bk_osc} induces oscillations in the non-linear squeezed matter bispectrum, which can be modeled non-perturbatively using the separate universe formalism. Below, we present a brief derivation of this model. For a detailed derivation of the non-linear squeezed matter bispectrum using the separate universe technique, see Refs.~\cite{Goldstein:2022hgr, Goldstein:2023brb, Goldstein:2024bky}. These works build on peak-background split derivations of scale-dependent bias in non-local primordial non-Gaussianity~\cite{Schmidt:2010gw, Scoccimarro:2011pz, Schmidt:2012ys, Desjacques:2016bnm, Cabass:2018roz}.

Consider a generic separable primordial squeezed bispectrum:
\begin{equation}\label{eq:b_sq_generic}
B_\Phi(\vq,\vk)=4 f_{\rm NL} F(q)G(k)P_{\Phi}(q)P_{\Phi}(k),
\end{equation}
where $F(q)$ and $G(k)$ are functions of the soft and hard modes, respectively, with $q\ll k$. At linear order in $f_{\rm NL}$, Eq.~\eqref{eq:b_sq_generic} is equivalent to a \emph{local} rescaling of the linear power spectrum by a potential dependent factor (see, \emph{e.g.},~\cite{Desjacques:2016bnm, Goldstein:2024bky}):
\begin{align}\label{eq:Pk_modification_general}
P_{m}^{\rm lin}(\vk|\vx)|_{\scriptscriptstyle\Phi_L(\vq)}&=\left[1+4f_{\rm NL}F(q)G(k)\,\Phi_L(\vq){ e^{\scriptstyle i\mathbf{q} \cdot \mathbf{x}}}\right]\nonumber\\
&\quad\,\times\,P_{m}^{\rm lin}(k),\nonumber\\
&\equiv \left[1+2\epsilon G(k)\right]P_{m}^{\rm lin}(k),
\end{align}
where $\epsilon\equiv 2f_{\rm NL}F(q)\Phi_L(\vq)$ includes all contributions from the long-wavelength background mode,\footnote{The factor of two ensures that the transformation is equivalent to the well-known rescaling of $\sigma_8\rightarrow \sigma_8(1+\epsilon)$ in local primordial non-Gaussianity ($F(q)=G(k)=1$).} which can be locally treated as a constant, and $\vx$ is the local position (which can be set to zero without loss of generality).

Since non-linear evolution is a (relatively) local process, this relation can be used to derive the impact of $f_{\rm NL}$ on late-time statistics. In particular, the squeezed bispectrum is given by the correlation of a long-wavelength mode, $\delta_{m,L}(\vq,z)$,\footnote{For clarity, we include all redshift dependencies here.} with the \textit{non-linear} power spectrum, $P_m(\vk,z|\vx)|_{\Phi_L}$, which itself can be written in terms of $\Phi_L$:
\begin{align*}\label{eq: squeezed_bispectrum_integral}
    \begin{split}
        B_m(\vq,\vk,z) &=\av{\delta_{ m,L}(\vq,z) P_{m}(\vk,z\,|\,\vx)|_{\Phi_L}}_c'\,, \\
        &=\av{\delta_{ m,L}(\vq)\left(\int_{\vp} \, \frac{\partial P_m(k)}{\partial\,\Phi_{L}(\vp)}\, \Phi_{L}(\vp) \right)}_c'\,, \\
        &=\frac{\partial P_m(\vk,z)}{\partial\,\Phi_{L}(\vq)}\frac{3 \Omega_mH_0^2}{2D_{\rm md}(z)}\frac{P_m(q,z)}{q^2T(q)} \,,\\
        &=\frac{3f_{\rm NL}\Omega_mH_0^2}{D_{\rm md}(z)}\frac{\partial P_m(k,z)}{\partial\,\epsilon}\bigg\vert_{\epsilon=0}\frac{F(q)P_m(q,z)}{q^2\,T(q)} \,,\\
    \end{split}
\end{align*}
where $\partial P_m(k,z)/\partial\epsilon\vert_{\epsilon=0}$ is the response of the non-linear matter power spectrum to a change in the linear power spectrum by Eq.~\eqref{eq:Pk_modification_general}. To derive this, we used the Poisson equation,

\begin{equation*}
\delta_{m,L}(\vq, z) = \frac{2 q^2 T(q) D_{\rm md}(z)}{3 \Omega_m H_0^2},
\end{equation*}
where $T(q)$ is the matter transfer function normalized to one on large scales, $D_{\rm md}(z)$ is the linear growth factor normalized to the scale factor during matter domination, $\Omega_m$ is the present-day matter density, and $H_0$ is the Hubble constant.

Crucially, if $F(q)\propto q^\Delta$ with $\Delta<2$, then the \emph{non-linear} squeezed bispectrum will include a pole $B_m(q,k)/P_m(q)\propto 1/q^{2-\Delta}$. Such a pole violates the LSS consistency condition~\cite{Kehagias:2013yd, Peloso:2013zw}, which states that, in the absence of primordial non-Gaussianity and equivalence principle violating physics, the ratio of the squeezed bispectrum to the long-wavelength power spectrum is protected from such poles.\footnote{The LSS consistency condition is particularly powerful because it is non-perturbative and thus remains valid in the non-linear regime. Moreover, there has been significant theoretical work generalizing and clarifying the LSS consistency relation~\cite{Creminelli:2004yq,Cheung:2007sv,Tanaka:2011aj,Creminelli:2012ed,Hinterbichler:2012nm,Assassi:2012zq,Kehagias:2012pd,Pajer:2013ana,Hinterbichler:2013dpa,Goldberger:2013rsa,Baldauf:2015xfa,Bravo:2017gct,Hui:2018cag}. The specific application of the consistency relation relevant for this analysis is discussed in detail in Refs.~\cite{Goldstein:2022hgr, Goldstein:2024bky}.} In this case, the leading-order $f_{\rm NL}$ contribution to the bispectrum can be cleanly separated from \emph{any} complicated non-linear contributions to the bispectrum that satisfy the equivalence principle, even in the non-linear regime (see Refs.~\cite{Goldstein:2022hgr, Giri:2023mpg, Goldstein:2024bky} for more details and validation of this point). Conversely, if $\Delta \geq 2$, such as in equilateral non-Gaussianity, we cannot cleanly separate the gravitational and primordial contributions to the bispectrum using the method presented here.

Applying this reasoning to the oscillatory bispectrum specified by Eq.~\eqref{eq:squeezed_bk_osc}, the non-linear squeezed matter bispectrum can be written as\footnote{We write all redshift dependencies in this expression for clarity.} 
\begin{widetext}
\begin{equation}\label{eq:B_squeezed_full}
\begin{split}
    \lim_{q\ll k}B_{m}(\vq,\vk,z)=&\frac{3f_{\rm NL}\Omega_mH_0^2}{D_{\rm md}(z)}\frac{P_m(q,z)}{q^{2-\Delta}T(q)}\bigg[\cos(\mu\ln q+\delta) \frac{\partial P_m(k,z|\mu,\Delta)}{\partial \epsilon_c}\bigg\vert_{\epsilon_c=0}+\sin(\mu\ln q+\delta) \frac{\partial P_m(k,z|\mu,\Delta)}{\partial \epsilon_s}\bigg\vert_{\epsilon_s=0}\bigg] \\
    & +\left(a_0(k)+{a}_2(k)\frac{q^2}{k^2}+\cdots\right)P_m(q,z)P_m(k,z).
\end{split}
\end{equation}
\end{widetext}
Here, the only non-trivial ingredients are the separate universe power spectrum derivatives: these can be estimated by running (Gaussian) simulations with the following transformed initial power spectra:
\begin{align}
    P^{\rm lin}_m(k) &\rightarrow  \left[ 1\pm 2\epsilon_c\,k^{-\Delta}\,\cos(\mu\ln k )\right]P^{\rm lin}_m(k),\label{eq:SU_transformation_cos}\\
    P^{\rm lin}_m(k) &\rightarrow  \left[ 1\pm 2\epsilon_s\,k^{-\Delta}\,\sin(\mu\ln k )\right]P^{\rm lin}_m(k),\label{eq:SU_transformation_sin} 
\end{align}
and computing finite differences.

We make a few comments on this result. First, note that Eq.~\eqref{eq:B_squeezed_full} closely resembles the cross-power spectrum between the matter density field and a biased tracer, with $a_0(k)$ and $a_2(k)$ playing the role of standard bias coefficients and ${\partial P_m(k|\mu,\Delta)}/{\partial \epsilon_c}\vert_{\epsilon_c=0}$ and ${\partial P_m(k|\mu,\Delta)}/{\partial \epsilon_s}\vert_{\epsilon_s=0}$ acting as non-Gaussian bias parameters that appear in cosmologies with primordial non-Gaussianity. This similarity arises because the squeezed bispectrum effectively describes the cross-power between a long-wavelength density mode and the locally measured small-scale power spectrum, which itself is just a biased tracer of $\delta_m(\vk)$~\cite{Giri:2023mpg}. Note that there are two response functions associated with this scenario, which arise from expressing the cosine term in a separable form and are analogous to the two non-Gaussian bias appearing in the Cosmological Collider scenario~\cite{Cabass:2024wob,Cabass:2018roz}. 

Second, the bottom line of Eq.~\eqref{eq:B_squeezed_full} represents the gravitational non-Gaussianity contribution to the squeezed matter bispectrum. We model this using the response approach~\cite{Valageas:2016hhr, Wagner:2014aka, Chiang:2014oga, Wagner:2015gva, Chiang:2017vsq, Esposito:2019jkb, Barreira:2017sqa, Biagetti:2022ckz}, where $a_0(k)$ and $a_2(k)$ are response functions that capture the leading-order terms allowed in the squeezed limit, with the $q$-dependence fixed by the equivalence principle. In principle, these coefficients can be derived from simulations (\emph{e.g.},~\cite{Chiang:2017jnm, Biagetti:2022ckz}) or from perturbation theory (\emph{e.g.},~\cite{Valageas:2013zda, Nishimichi:2014jna}). However, following Refs.~\cite{Esposito:2019jkb, Goldstein:2022hgr}, we treat them as free parameters and marginalize over them in our analysis. In practice, as described in Sec.~\ref{subsec:measurements_modelling_likelihood}, we measure the squeezed bispectrum averaged over modes with $k_{\rm min}<k,k'<k_{\rm max}$; hence the response functions reduce to scalars $\bar{a}_0^{(i)}$ and $\bar{a}_2^{(i)}$ for each $k$ bin. 

\begin{table}[!t]
    \centering

    \begin{tabular}{ |c|c|c|c| } 
    
     \multicolumn{4}{c}{{\centering{\it PNG Simulations}}}          \\
     \hline
     \hline
   $~~f_{\rm NL}~~$ & $~\mu~$ & $~~\delta~~$  & $N_{\rm sim}$ \\ 
    \hline
    +300  & 4  &  $3\pi/4$ & 35  \\
    \hline
    -300  & 4  &  $3\pi/4$ & 35  \\
    \hline
    \end{tabular}
    \vspace{15pt}
    \begin{tabular}{ |c|c|c|c| } 
         \multicolumn{4}{c}{~~\it{Separate Universe}~~}          \\
     \hline
     \hline
   $~~\pm \epsilon_c~~$ & $~~\pm \epsilon_s~~$  & $\mu$ & $N_{\rm sim}$ \\ 
    \hline
    0.015 &  0 &  4 & 3  \\
    \hline
     0 &  0.015  &  4 & 3  \\
     \hline
    \end{tabular}
    
   \caption{Parameter values for the simulations used in this work. \emph{Left}: settings for simulations including the squeezed primordial bispectrum in Eq.~\eqref{eq:squeezed_bk_osc}. We run 35 realizations for each value of $f_{\rm NL}$, resulting in 70 total simulations with oscillatory bispectra.
   \emph{Right}: settings for the separate universe simulations. We run three realizations each with the linear power spectrum rescaled according to Eq.~\eqref{eq:SU_transformation_cos} and Eq.~\eqref{eq:SU_transformation_sin}, assuming $\pm\epsilon_c=0.015$ and $\pm\epsilon_s=0.015$, respectively. For each separate universe simulation, we also run a corresponding paired-and-fixed simulation to reduce cosmic variance when computing the response functions, resulting in a total of 24 separate universe simulations.}
   \label{tab:sim_parameters}
\end{table}

\section{Methodology}\label{Sec:methodology}
\subsection{Simulations}

\noindent In this section, we describe the two types of simulations used in this work. The first set of simulations have modified initial conditions that include the primordial bispectrum specified by Eq.~\eqref{eq:squeezed_bk_osc}. The second set are separate universe simulations with Gaussian initial conditions, but modified power spectra. We use these separate universe simulations to compute the potential derivatives, $\partial P_m(k)/\partial \epsilon$, in Eq.~\eqref{eq:B_squeezed_full}. Table~\ref{tab:sim_parameters} summarizes the key parameters for the simulations used in this work. Note that we fix $\Delta=0$ for all simulations used in the main text.

Aside from the modified initial conditions (discussed below), all simulations have the same settings and seeds as the corresponding fiducial runs in the \textsc{Quijote} suite~\cite{Villaescusa-Navarro:2019bje, Coulton:2022rir}.\footnote{\href{https://quijote-simulations.readthedocs.io/en/latest/}{https://quijote-simulations.readthedocs.io/en/latest/}} Specifically, each simulation contains $512^3$ particles in a $(1~{\rm Gpc}/h)^3$ comoving volume, with $\Omega_m=0.3175$, $\Omega_{b}=0.049$, $h=0.6711$, $n_s=0.9624$, and $\sigma_8=0.834$. The simulations were run using the \textsc{Gadget-3} treePM code~\cite{Springel:2005mi}. We use the density fields at redshifts $z=3,$ $1$, and $0.$ For further details on the \textsc{Quijote} simulations, see Ref.~\cite{Villaescusa-Navarro:2019bje}.

\subsubsection{Initial conditions with oscillatory bispectra}\label{subsec:PNG_IC_sims}

\noindent Here, we discuss the procedure used to generate non-Gaussian initial conditions with the oscillatory squeezed bispectrum in Eq.~\eqref{eq:squeezed_bk_osc}. We utilize the algorithm derived in Appendix A of~\citet{Goldstein:2024bky}, which is based off of the procedure for generating non-Gaussian initial conditions with a generic bispectrum from Ref.~\cite{Scoccimarro:2011pz}.

Starting from a Gaussian primordial potential, $\phi_G(\vk)$, we construct a non-Gaussian potential
\begin{equation}\label{eq:quadratic_png}
    \Phi(\vk)=\phi_G(\vk)+f_{\rm NL}\left[\Psi(\vk) -\langle \Psi(\vk)\rangle\right],
\end{equation}
where $f_{\rm NL}$ is the amplitude of non-Gaussianity, and $\Psi(\vk)$ is an auxiliary field that is quadratic in $\phi_G(\vk)$, \emph{i.e.}, 
\begin{equation*}
    \Psi(\vk)=\int\limits_{\vk_1,\vk_2}(2\pi)^3\delta_D(\vk-\vk_{12})\,K(\vk_1, \vk_2)\,\phi_G(\vk_1)\,\phi_G(\vk_2).
\end{equation*}
Here, $\vk_{12}\equiv \vk_1+\vk_2$ and $K(\vk_1, \vk_2)$ is a kernel that is chosen\footnote{For a given bispectrum, the coupling kernel is not uniquely determined~\cite{Schmidt:2010gw, Scoccimarro:2011pz}. It is important to choose $K(\vk_1,\vk_2)$ such that loop contributions do not alter the $k^{n_s-4}$ scaling of the large-scale power spectrum.} to ensure that the squeezed bispectrum of $\Phi$ matches Eq.~\eqref{eq:squeezed_bk_osc} at leading order in $f_{\rm NL}$. One choice for this kernel is
\begin{widetext}
\begin{align}\label{eq:IC_kernel}
\begin{split}
    K(\vk_1,\vk_2)&=\left\{\left(\frac{k_1}{k_{12}}\right)^\Delta\cos\left(\mu\ln\left(\frac{k_1}{k_{12}}\right)+\delta\right)+ \left(\frac{k_2}{k_{12}}\right)^\Delta\cos\left(\mu\ln\left(\frac{k_2}{k_{12}}\right)+\delta\right)-\cos(\delta) \right\},\\
    &=\frac{1}{2}\left\{ e^{i\delta}\left[\left(\frac{k_1}{k_{12}}\right)^{\Delta+i\mu}+\left(\frac{k_2}{k_{12}}\right)^{\Delta+i\mu}-1 \right]+e^{-i\delta}\left[\left(\frac{k_1}{k_{12}}\right)^{\Delta-i\mu}+\left(\frac{k_2}{k_{12}}\right)^{\Delta-i\mu}-1 \right]\right\},
\end{split}
\end{align}
\end{widetext}
where the second line is written in a separable form that can be easily evaluated using FFTs.\footnote{The separable kernel includes complex terms, thus this kernel needs to be implemented with care if using an initial condition code based on real FFTs.} Assuming a scale-invariant power spectrum, Eq.~\eqref{eq:IC_kernel} produces the target squeezed bispectrum as long as $\Delta \ll 3/2$, which is the case for the simulations used here. The kernel can be generalized to simulate larger values of $\Delta$ by multiplying by a high-pass filter~\cite{Goldstein:2024bky}.

We generate initial conditions with Eq.~\eqref{eq:IC_kernel} using a modified version of \texttt{2LPTPNG}.\footnote{\href{https://cosmo.nyu.edu/roman/2LPT/}{https://cosmo.nyu.edu/roman/2LPT/}} We run 35 oscillatory bispectrum simulations each with $f_{\rm NL}=+300$ and $f_{\rm NL}=-300$. We fix $\Delta=0$, $\mu=4$ and $\delta=3\pi/4$. Our choice of $\mu$ is motivated by two considerations: it is large enough to resolve oscillations across the scales of interest ($0.01 \lesssim k \lesssim 1~h/{\rm Mpc}$), yet small enough to remain observationally relevant, given that higher frequencies are Boltzmann suppressed in the Cosmological Collider scenario. Finally, we fix the phase $\delta=3\pi/4$ to provide a non-trivial test of our initial condition generation and parameter estimation pipelines. 

For comparison, we also analyze 35 realizations of the fiducial (Gaussian) \textsc{Quijote} simulations and the \texttt{LC\_p}  ($f_{\rm NL}^{\rm loc}=+100$) \textsc{Quijote-PNG}, using the same random seeds as our oscillatory bispectrum simulations. We use the simulations with Gaussian initial conditions to isolate the impact of of primordial non-Gaussianity from gravitational non-Gaussianity. We use simulations with local primordial non-Gaussianity to compare how well reconstruction improves constraints on primordial non-Gaussianity \emph{without} oscillations.

\subsubsection{Separate Universe Simulations}\label{subsec:separate_universe}

\noindent We run separate universe simulations to compute the sine and cosine potential derivatives appearing in Eq.~\eqref{eq:B_squeezed_full}. Specifically, we rescale the fiducial power spectrum for the cosine and sine simulations using Eq.~\eqref{eq:SU_transformation_cos} and Eq.~\eqref{eq:SU_transformation_sin}, respectively. We choose $\epsilon_c=\epsilon_s=\pm 0.015$, which is sufficient to numerically estimate the response functions. To suppress cosmic variance, we run six realizations for each value of $\epsilon_c$ and $\epsilon_s$: three realizations with independent random seeds and three with paired-and-fixed initial conditions (identical seeds but inverted phases)~\cite{Pontzen:2015eoh, Angulo:2016hjd,Villaescusa-Navarro:2018bpd}. This yields a total of 24 separate universe simulations -- 6 simulations for each value of $\pm \epsilon_c$ and $\pm \epsilon_s$. To compute the response functions for local primordial non-Gaussianity, we use six realizations each from the \texttt{s8\_m} ($\sigma_8 = 0.819$) and \texttt{s8\_p} ($\sigma_8=0.849$) runs in the \textsc{Quijote} suite.

\subsection{Iterative reconstruction}
\noindent Previous studies have shown that oscillatory features in the power spectrum, which are predominately damped by large-scale bulk flows, can be partially restored by applying a non-linear operation on the density field via a procedure known as reconstruction. Broadly speaking, reconstruction algorithms estimate the large-scale displacement field, typically by applying perturbation theory to a smoothed version of the non-linear density field, and then shift the density field by the negative of this displacement to estimate the linear density field~\cite{Eisenstein:2006nk, Padmanabhan:2008dd, Schmittfull:2017uhh,Chen:2024eri}. 
 
While a detailed analysis of reconstruction algorithms in the context of oscillatory bispectra is beyond the scope of this work, we present a preliminary investigation into how well reconstruction can restore oscillations in the squeezed bispectrum. To this end, we use the iterative reconstruction algorithm developed in \citet{Schmittfull:2017uhh} and implemented in the publicly available \texttt{iterrec} code.\footnote{\href{https://github.com/mschmittfull/iterrec}{https://github.com/mschmittfull/iterrec}}. This algorithm estimates the displacement field by iteratively moving (matter) particles according to their Zel’dovich displacement estimated from a smoothed density field, with a decreasing smoothing scale at each step.  The total displacement is then estimated by comparing the initial particle positions with their positions after the final iteration. The reconstructed linear density field is obtained from the divergence of the total displacement field.\footnote{This corresponds to the $\mathcal{O}(1)$ iterative reconstruction method described in Ref.~\cite{Schmittfull:2017uhh}.} This iterative approach has been shown to outperform conventional reconstruction techniques, particularly in recovering the linear matter density field.

In our analysis, we perform iterative reconstruction on a $512^3$ mesh with cloud-in-cell (CIC) interpolation. We use four iterations with Gaussian smoothing scales between $R_{\rm ini.}=20~{\rm Mpc}/h$ and $R_{\rm fin.}=2.5~{\rm Mpc}/h$, halving the smoothing scale after each iteration. We reconstruct the redshift $z=0$ snapshots for all 35 realizations of the $f_{\rm NL}=+300$ oscillatory bispectrum simulations, as well as the corresponding 35 realizations with Gaussian initial conditions and those with $f_{\rm NL}^{\rm loc}=+100$.

\subsection{Measurements and likelihood}\label{subsec:measurements_modelling_likelihood}

\noindent In this section, we describe our measurements and likelihood. We divide our analysis into a ``fine-binned analysis," used to qualitatively study the non-linear evolution of oscillatory features in the bispectrum, and a ``coarse-binned analysis," used to quantitatively validate our non-perturbative bispectrum model (Eq.~\ref{eq:B_squeezed_full}). The fine-binned analysis is used purely for visual purposes, allowing us to compare the non-linear evolution of the oscillations in the bispectrum with predictions based on perturbation theory and separate universe simulations. In contrast, the coarse-binned analysis, which uses significantly wider hard mode bins, is designed to quantify how well we can constrain $f_{\rm NL}$ and $\delta$ using the squeezed matter bispectrum. This coarser binning is essential for reliably estimating the covariance matrix given the 35 realizations used in this work. Before presenting the specific details of each analysis, we outline the choices common to both.

Our methodology closely follows the bispectrum analysis described in~\citet{Goldstein:2024bky}, to which we refer the reader for additional details. We assign particles to a $1024^3$ mesh using CIC interpolation with the \textsc{Pylians} package~\cite{Pylians}.\footnote{\href{https://pylians3.readthedocs.io/en/master/}{https://pylians3.readthedocs.io/en/master/}} We use the \texttt{PNGolin} package\footnote{\href{https://github.com/samgolds/PNGolin}{https://github.com/samgolds/PNGolin}} to compute all spectra and binning corrections. We define bins in units of the fundamental mode, $k_f\equiv 2\pi/L_{\rm box}\approx 0.006~h/{\rm Mpc}$. Since our squeezed bispectrum measurements include some of the longest wavelength modes accessible in our simulation volume, it is important to account for the finite number of modes available at low $q$ when computing theoretical predictions. To this end, we compute $\langle F(q)\delta_m(\vq) \delta_m(-\vq)\rangle$ directly from the density fields, where $F(q)\in\{ \sin(\mu\ln q)/(q^2T(q)),\,   \cos(\mu\ln q)/(q^2T(q)),\,  q^2 \}$. Similarly, to account for both binning corrections when averaging over hard modes and the influence of $f_{\rm NL}$ on the small-scale matter power spectrum, we compute $\langle G(k) \delta_m(\vk) \delta_m(-\vk)\rangle$ on the grid. Here, $G(k)\in\{\partial \ln P_m(k|\mu)/{\partial \epsilon_c}\vert_{\epsilon_c=0}, \partial P_m(k|\mu)/{\partial \epsilon_s}\vert_{\epsilon_s=0}\}$ is the separate universe response function, and $\delta_m(\vk)$ is estimated from the particular simulation that we are analyzing.

To reduce cosmic variance, we analyze the mean data-vector averaged over 35 realizations, with the covariance estimated from the same set of simulations and rescaled by a factor of $1/35$. This corresponds to an effective volume of $V_{\rm eff}=35~({\rm Gpc}/h)^3$, which exceeds that of current surveys like DESI~\cite{DESI:2016fyo, Ding:2022ydj} and allows for stringent tests of our theoretical modeling. We assume that the bispectrum covariance is diagonal in the soft mode $q$, which we always take to be in the linear regime, but includes off-diagonal correlations in $k$ to include the significant coupling between hard modes in the squeezed bispectrum~\cite{Biagetti:2021tua}. For post-reconstruction analyses, we adopt the same pipeline as pre-reconstruction, except we use a $512^3$ mesh to match resolution of the reconstruction. This has a negligible impact on our results as we only fit the post-reconstruction bispectrum up to $k_{\rm max} = 0.5~h/{\rm Mpc}$, whereas the Nyquist frequencies for the $1024^3$ and $512^3$ grids are $3.2~h/{\rm Mpc}$ and $1.6~h/{\rm Mpc}$, respectively.

\subsubsection{Fine-binned analysis}

\noindent To qualitatively study oscillatory features in the late-time bispectrum, we compute the squeezed bispectrum with soft modes in four linearly spaced bins between $0.006\approx 2k_f\leq q<6 k_f\approx0.038~h/{\rm Mpc}$ and hard modes in 17 logarithmically spaced bins between $0.005\approx 8k_f<k<150k_f\approx 0.94~{h/{\rm Mpc}}$. We also compute the matter power spectrum and separate universe response functions in their corresponding $q$ and $k$ bins. From these measurements, we computing the following statistic
\begin{equation}\label{eq:Bspec_differnce}
    \frac{\Delta B(q,k)}{P(q)P(k)}\equiv \frac{B^{f_{\rm NL}^\pm}_m(q,k)}{P^{f_{\rm NL}^\pm}_m(q)P^{f_{\rm NL}^\pm}_m(k)}-\frac{B^{G}_m(q,k)}{P^G_m(q)P^{G}_m(k)},
\end{equation}
where $B^{f_{\rm NL}^\pm}$ and $P^{f_{\rm NL}^\pm}$ are estimated from the simulations with $f_{\rm NL}=\pm 300$, and $B^{G}$ and $P^{G}$ are estimated from simulations with Gaussian initial conditions.\footnote{Alternatively, one could take the difference of simulations with positive and negative $f_{\rm NL}$ to cancel the $f_{\rm NL}^2$ contributions to the bispectrum. We compute this difference in Appendix~\ref{App:O_fnlsq_contribution} and find that these terms are negligible for the scale cuts, simulation volume, and redshifts analyzed here.} 

Eq.~\eqref{eq:Bspec_differnce} is a useful statistic for qualitatively analyzing the non-linear evolution of oscillatory features in the squeezed bispectrum for several reasons. First, it cancels the otherwise dominant contribution to the bispectrum coming from gravitational non-Gaussianity. Second, it largely cancels sample variance from the finite number of soft modes. Finally, at tree-level, Eq.~\eqref{eq:Bspec_differnce} takes a particularly simple form --- it is directly proportional to $\cos(\mu\ln(q/k)+\delta)$. 

Despite these benefits, measurements of Eq.~\eqref{eq:Bspec_differnce} cannot be directly compared to the separate universe term in the top line of Eq.~\eqref{eq:B_squeezed_full} because Eq.~\eqref{eq:Bspec_differnce} includes subdominant contributions coming from the coupling of primordial non-Gaussianity to gravity. These terms are naturally accounted for in our theory model (Eq.~\ref{eq:B_squeezed_full}) via small shifts in $\bar{a}_0^{(i)}$ and $\bar{a}_2^{(i)}$ when fitting measurements with and without primordial non-Gaussianity. For visualization purposes when comparing measurements of Eq.~\eqref{eq:Bspec_differnce} with the separate universe response functions, we subtract off this small contribution using a procedure described in Appendix~\ref{App:O_fnlsq_contribution}.

\begin{figure*}[!t]
\centering
\includegraphics[width=0.995\linewidth]{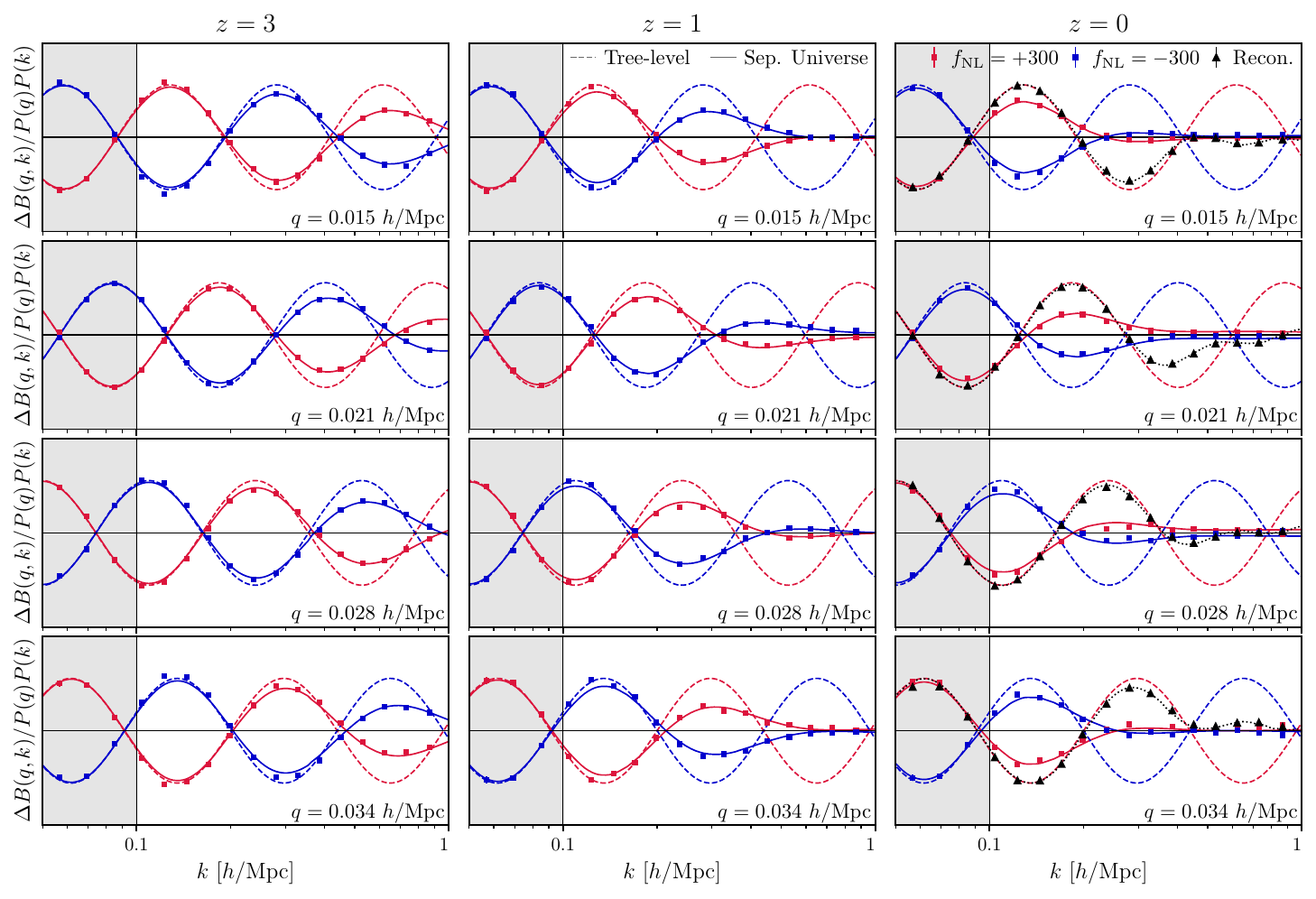}
\caption{Non-linear evolution of oscillations in the squeezed bispectrum across redshifts and scales. To isolate the primordial contribution, we compute the difference in the normalized bispectrum $\Delta B(q,k)/P(q)P(k)$ between simulations with and without $f_{\rm NL}$. The horizontal (vertical) panels show different redshifts (soft modes). Notably, the oscillations are significantly damped by non-linear evolution, though this is accurately described by our separate universe model (solid lines). The black points show the same measurement estimated after applying iterative reconstruction to the density field, which significantly reduces the damping. The post-reconstruction bispectrum  agrees remarkably well with theoretical predictions using the reconstructed separate universe simulations. The gray shaded regions denote triangle configurations that are not sufficiently squeezed, hence our non-perturbative model is not expected to be valid there. Nevertheless, as discussed in the text, the bispectrum in these regions is still dominated by squeezed contributions due to the way the initial conditions were generated, thus our model still agrees with the measurements in these configurations.}
\label{fig:full_bispectrum_meas_Delta0}
\end{figure*}
\subsubsection{Coarse-binned analysis}

\noindent In order to demonstrate that our non-perturbative bispectrum model (Eq.~\ref{eq:B_squeezed_full}) can be used to obtain unbiased constraints on the amplitude ($f_{\rm NL}$) and phase ($\delta$), we estimate the squeezed bispectrum integrated over a range of hard modes. Specifically, we compute the bispectrum in six soft mode bins linearly spaced between $0.006~h/{\rm Mpc}\approx k_f\leq q < 7k_f \approx 0.044~h/{\rm Mpc}$ and four hard mode bins logarithmically spaced between $20k_f\approx0.13~h/{\rm Mpc}<k<k_{\rm max},$ where we consider three different values of $k_{\rm max}\in [50k_f, 80k_f,160k_f]\approx [0.3, 0.5, 1.0]~h/{\rm Mpc}$. 

We perform a joint analysis of the squeezed matter bispectrum $B_m(q,k)$ and the long-wavelength matter power spectrum $P_m(q)$, varying $f_{\rm NL}$, $\delta$, $\bar{a}_0^{(i)}$, and $\bar{a}_2^{(i)}$ (with $1 \leq i \leq 4$ labeling the hard mode bins). Although a complete analysis would additionally vary the frequency $\mu$, we fix the frequency to its true value to avoid the need to run additional separate universe simulations. Following~\citet{Goldstein:2022hgr}, we express the joint likelihood for $B_m(q,k)$ and $P_m(q)$ as a likelihood in $B_m(q,k)$ with a parameter-dependent covariance and a theory model that depends on the \emph{measured} $P_m(q)$,\footnote{We use a multivariate $t$-distribution likelihood since we estimate the covariance from a finite number of simulations~\cite{Sellentin:2015waz}.}
\begin{align} \label{eq:likelihood}
    \mathcal{L} \propto \frac{1}{\sqrt{\det {\mathcal{C}}({\bm{\theta}})}}\bigg[1+\frac{\delta B({\bm{\theta}})\cdot{\mathcal{C}}^{-1}({\bm{\theta}})\cdot\delta  B({\bm{\theta}})}{N_{\rm sim}-1}\bigg]^{-\frac{N_{\rm sim}}{2}} \,,
\end{align}
where $\bm{\theta}$ denotes the parameter vector, $\delta B({\bm{\theta}}) \equiv {B}_m(q,k)-B_{\rm thr.}(q,k|{\bm{\theta}})$ is the difference between the measured and theory bispectrum, 
${\mathcal{C}}_{ij}({\bm{\theta}})\equiv \langle\delta B_i({\bm{\theta}})\delta B_j({\bm{\theta}})\rangle$ is the \emph{parameter-dependent} covariance of $\delta B$, and $N_{\rm sim}=35$ is the number of realizations used to estimate the covariance. Crucially, $B_{\rm thr.}(q,k|{\bm{\theta}})$ is computed using Eq.~\eqref{eq:B_squeezed_full} with $a_0(k)$ and $a_2(k)$ replaced by $\bar{a}_0^{(i)}$ and $\bar{a}_2^{(i)}$, and with all occurrences of $P_m(q,z)F(q)$ replaced by $\langle F(q)\delta_m(\vq) \delta_m(-\vq)\rangle$ computed from the particular realization, where $F(q)\in\{ \sin(\mu\ln q)/(q^2T(q)),\,   \cos(\mu\ln q)/(q^2T(q)),\,  q^2 \}$. This procedure is equivalent to a joint likelihood in the bispectrum and power spectrum and refer the reader to Ref.~\cite{Goldstein:2022hgr} for more details about this point.

 We sample the posterior using the \texttt{emcee} package~\cite{Foreman-Mackey:2012any}, assuming wide uniform priors: $-10^4\leq f_{\rm NL}\leq 10^4$, $-20\leq \bar{a}_0^{(i)}, \, \bar{a}_2^{(j)}\leq 20$ and $0 \leq \delta < 2\pi$.  Directly sampling the posterior including the parameter-dependent covariance is computationally expensive. Consequently, we fix the covariance to its value at the MAP parameters, determined via the iterative maximization procedure outlined in Ref.\cite{Goldstein:2022hgr}. This approximation was shown in Ref.\cite{Goldstein:2022hgr} to have a negligible impact on the final results. We emphasize that this MAP-based covariance differs from the usual bispectrum covariance, which corresponds to the parameter-dependent covariance with all parameters fixed to zero.

\begin{figure}[!t]
\centering
\includegraphics[width=0.99\linewidth]{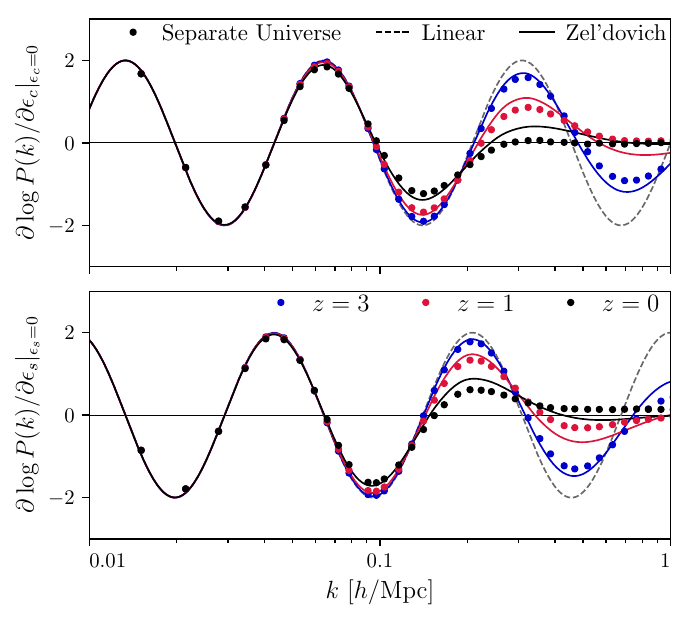}
\caption{Non-linear contributions to the response function -- the central ingredient of our late-time squeezed bispectrum model. The solid points show non-perturbative estimates of the response of the matter power spectrum to a local rescaling of the linear power spectrum by Eq.~\eqref{eq:SU_transformation_cos} (top) and~\eqref{eq:SU_transformation_sin} (bottom) estimated from separate universe simulations. The dashed (solid) lines show predictions based using linear theory (the Zel'dovich approximation), with different colors denoting different redshifts. The damping is significant at low redshifts, but accurately described by the Zel'dovich approximation in the quasi-linear regime.}
\label{fig:response_function}
\end{figure}
\section{Results}\label{Sec:results}

\noindent Fig.~\ref{fig:full_bispectrum_meas_Delta0} illustrates the impact of oscillations in the primordial squeezed bispectrum on the late-time squeezed matter bispectrum. Specifically, we show the difference $\Delta B(q,k)/P(q)P(k)$ between simulations with and without primordial non-Gaussianity defined in Eq.~\eqref{eq:Bspec_differnce}, across a range of redshifts and for $f_{\rm NL}=\pm300$. At large scales and high redshifts, the measured bispectrum is consistent with tree-level perturbation theory (dashed lines in the figure); however, at small scales and low redshifts, the squeezed bispectrum is significantly damped. Notably, the separate universe model given in Eq.~\eqref{eq:B_squeezed_full} (solid lines in the figure) provides an excellent description of the damping observed in our bispectrum simulations across all scales and redshifts analyzed, validating our approach. 

We additionally plot contributions with $k\lesssim 0.1~h/{\rm Mpc}$ (gray bands), which are not necessarily in the squeezed limit for all $q$ bins. Although our model is formally only valid in the squeezed limit, we still find good agreement between theory and measurement in these semi-squeezed regimes. This is a consequence of our initial condition setup, which generates a primordial bispectrum that is dominated by its squeezed-limit contribution -- even for triangles that are only moderately squeezed. Indeed, at tree level, the largest difference between the squeezed bispectrum and full bispectrum template for the range of scales shown in Fig.~\ref{fig:full_bispectrum_meas_Delta0} is only 10\%. 

In the right panel of Fig.~\ref{fig:full_bispectrum_meas_Delta0}, we also plot the post-reconstruction bispectrum, which is estimated by taking the difference of the measurements from the reconstructed density field from the $f_{\rm NL}=+300$ simulations and the reconstructed density field from the simulations with Gaussian initial conditions. Iterative reconstruction largely undoes the damping up to $k\approx 0.5~h/{\rm Mpc}$, substantially enhancing the oscillatory signal compared to the pre-reconstruction data. Furthermore, the post-reconstruction bispectrum is well-described by the same non-perturbative model used in the pre-reconstruction case, albeit with potential derivatives replaced by those estimated from the post-reconstruction separate universe simulations. In other words, our non-perturbative model can be immediately generalized to post-reconstruction fields by simply running the reconstruction pipeline on the separate universe simulations.\footnote{To make this more rigorous, let $\tilde{\delta}_m(\vk)$ denote the reconstructed density field. Since the pre- and post-reconstruction fields are equivalent at linear order, the post-reconstruction squeezed bispectrum is $\tilde{B}_m(\vq,\vk)=\langle \tilde{\delta}_m(\vq) \tilde{P}_m(\vk,z|\vx)\rangle_c'=\langle {\delta}_m(\vq) \tilde{P}_m(\vk,z|\vx)\rangle_c'$, which can be expanded as potential derivatives of the post-reconstruction power spectrum using the procedure in Sec.~\ref{Sec:background}.} For $k\gtrsim0.5~h/{\rm Mpc}$, the post-reconstruction bispectrum is broadly consistent with zero, with small departures that likely arise from imperfect reconstruction of small-scale non-linearities. More sophisticated reconstruction algorithms that try to reconstruct small-scale features could potentially restore the oscillations at even smaller scales~\cite{Parker:2025mtg}.

We stress that, aside from a small residual contribution from the coupling of primordial non-Gaussianity to gravity, which we subtract from the measurements in this plot (see Appendix~\ref{App:O_fnlsq_contribution}), the separate universe predictions for the pre- and post-reconstruction bispectrum shown in Fig.~\ref{fig:full_bispectrum_meas_Delta0} are not fits to the data. In conclusion, our non-perturbative separate universe model (Eq.~\ref{eq:B_squeezed_full}) provides an excellent description of the non-linear evolution of oscillations in the squeezed matter bispectrum, both before and after reconstruction.

\begin{figure}[!t]
\centering
\includegraphics[width=0.99\linewidth]{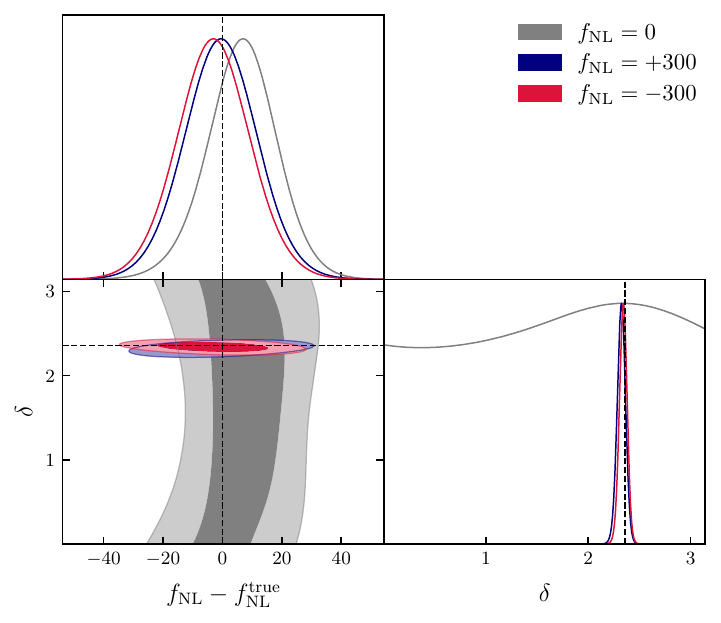}
\caption{Constraints on $f_{\rm NL}$ and $\delta$ from the squeezed matter bispectrum using simulations with three different fiducial values of $f_{\rm NL}$ at redshift $z=1$, assuming $k_{\rm max}=0.3~h/{\rm Mpc}.$ The dashed lines denote the true values of $f_{\rm NL}-f_{\rm NL}^{\rm true}$ and $\delta$. In all cases, we find precise recovery of the input amplitude, and, when $f_{\rm NL}^{\rm true}$ is non-zero, the phase.}
\label{fig:fNL_value_dependence}
\end{figure}

Fig.~\ref{fig:response_function} shows the separate universe response functions associated with Eq.~\eqref{eq:SU_transformation_cos} (top) and Eq.~\eqref{eq:SU_transformation_sin} (bottom). By redshift $z=0$, the oscillatory signal is almost completely erased for $k\gtrsim 0.3~h/{\rm Mpc}.$ For comparison, we also show the response functions estimated with the Zel'dovich approximation computed using \texttt{velocileptors}.\footnote{\href{https://github.com/sfschen/velocileptors}{https://github.com/sfschen/velocileptors}} The Zel'dovich approximation provides a good approximation of the damping in the quasi-linear regime, suggesting that the damping is primarily driven by long-wavelength displacements. These findings are consistent with previous numerical analyses of primordial features in the power spectrum~\cite{Vlah:2015zda, Ballardini:2019tuc, Chen:2020ckc, Euclid:2023shr, Calderon:2025xod, Stahl:2025qru}.

Having qualitatively investigated the non-linear evolution of oscillations in the squeezed matter bispectrum, we now turn to a quantitative assessment of our bispectrum model -- Eq.~\eqref{eq:B_squeezed_full}. 
Fig.~\ref{fig:fNL_value_dependence} shows the marginalized posterior on $f_{\rm NL}-f_{\rm NL}^{\rm true}$ and $\delta$ from a joint analysis of the squeezed matter bispectrum and matter power spectrum at redshfit $z=1$ up to $k_{\rm max}=0.3~h/{\rm Mpc}$. We show the results from simulations with $f_{\rm NL}=\pm 300$, as well as Gaussian initial conditions -- in all scenarios, we recover unbiased constraints on $f_{\rm NL}$ and $\delta$. Moreover, even in the absence of a detection of $f_{\rm NL},$ we find that marginalizing over the phase does not significantly degrade the constraint on $f_{\rm NL}.$ Note that the small bump in $\delta$ for the simulations with Gaussian initial conditions is due to the fact that we only use four hard mode bins in this fit and the precise locations of these bins impacts the sensitivity to different phases. These tests demonstrate that we can recover unbiased constraints on $f_{\rm NL}$ and $\delta$ regardless of the value used in the simulation.

\begin{figure}[!t]
\centering
\includegraphics[width=0.99\linewidth]{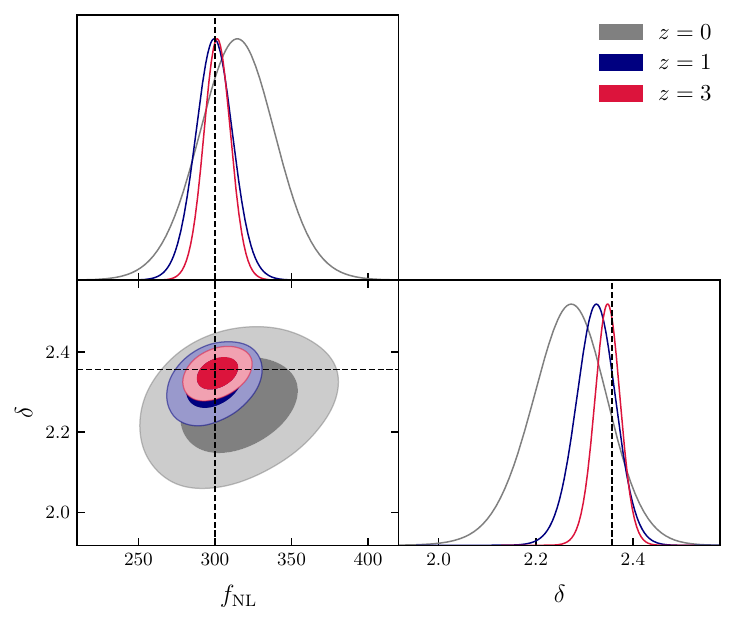}
\caption{Comparison of constraints on $f_{\rm NL}$ and $\delta$ from the squeezed matter bispectrum at three different redshifts for simulations with $f_{\rm NL}=300$, assuming $k_{\rm max}=0.3~h/{\rm Mpc}.$ The dashed lines indicate the true values of $f_{\rm NL}$ and $\delta$.}
\label{fig:fNL_redshift_dependence}
\end{figure}

Fig.~\ref{fig:fNL_redshift_dependence} shows the marginalized posterior on $f_{\rm NL}$ and $\delta$ from a joint analysis of the squeezed matter bispectrum and matter power spectrum using measurements from the $f_{\rm NL}=300$ simulations at three different redshifts. We recover unbiased constraints on $f_{\rm NL}$ and $\delta$ in all cases, indicating that our theory model accurately describes the redshift evolution of our simulations. Our constraint on $f_{\rm NL}$ is significantly tighter at higher redshifts, which is consistent with the damping seen in Fig.~\ref{fig:full_bispectrum_meas_Delta0}. For fixed $k_{\rm max}=0.3~h/{\rm Mpc}$, the 68\% error on $f_{\rm NL}$ ($\delta$) from the $z=0$ simulation is a factor of 2.9 (3.3) larger than the 68\% error from the $z=3$ simulations. As discussed in more detail below, we can find even better improvement at high redshifts by increasing $k_{\rm max}.$ This implies that high-redshift data will be invaluable for future searches of oscillatory features in the bispectrum.

\begin{figure}[!t]
\centering
\includegraphics[width=0.99\linewidth]{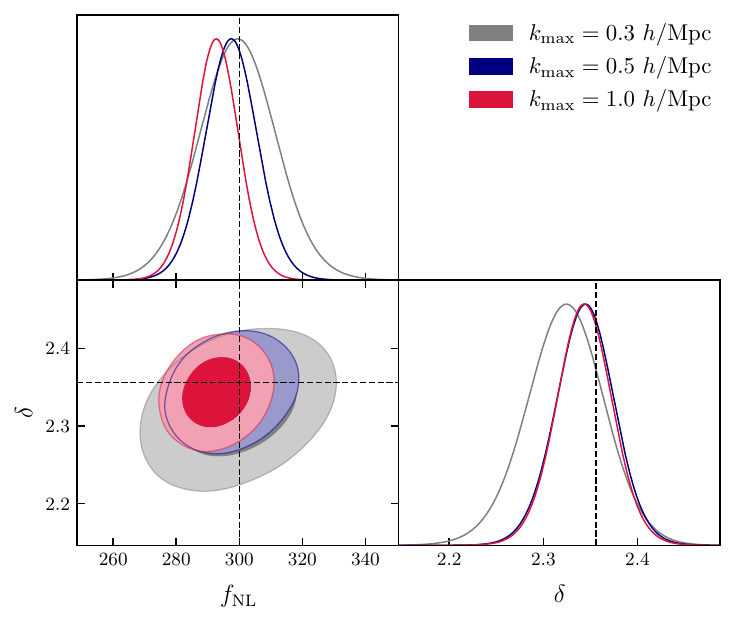}
\caption{Comparison of constraints on $f_{\rm NL}$ and $\phi$ from the squeezed matter bispectrum with three different values of the maximum hard mode, $k_{\rm max}$, at redshift $z=1.$ The dashed lines indicate the true values of $f_{\rm NL}$ and $\delta$. We recover unbiased constraints deep into the non-linear regime ($k_{\rm max}=1.0~h/{\rm Mpc}$), which is only possible because our squeezed bispectrum model is non-perturbative.}
\label{fig:fNL_scalecut_dependence}
\end{figure}

Fig.~\ref{fig:fNL_scalecut_dependence} shows the marginalized posterior on $f_{\rm NL}$ and $\delta$ from a joint analysis of the $z=1$ squeezed matter bispectrum and matter power spectrum as a function of the maximum hard mode, $k_{\rm max}$. We recover unbiased constraints on $f_{\rm NL}$ and $\delta$ for all scale-cuts up to $k_{\rm max}=1.0$, with the error on $f_{\rm NL}$ reducing by approximately 30\% (40\%) after extending $k_{\rm max}$ from $0.3~h/{\rm Mpc}$ to $0.5$ ($1.0~{h/{\rm Mpc}}$). Going beyond $k_{\rm max}=1.0~h/{\rm Mpc}$ has negligible improvements on the constraints at this redshift as the primordial oscillations are completely damped by this scale (see Fig.~\ref{fig:full_bispectrum_meas_Delta0}).

\begin{figure}[!t]
\centering
\includegraphics[width=0.99\linewidth]{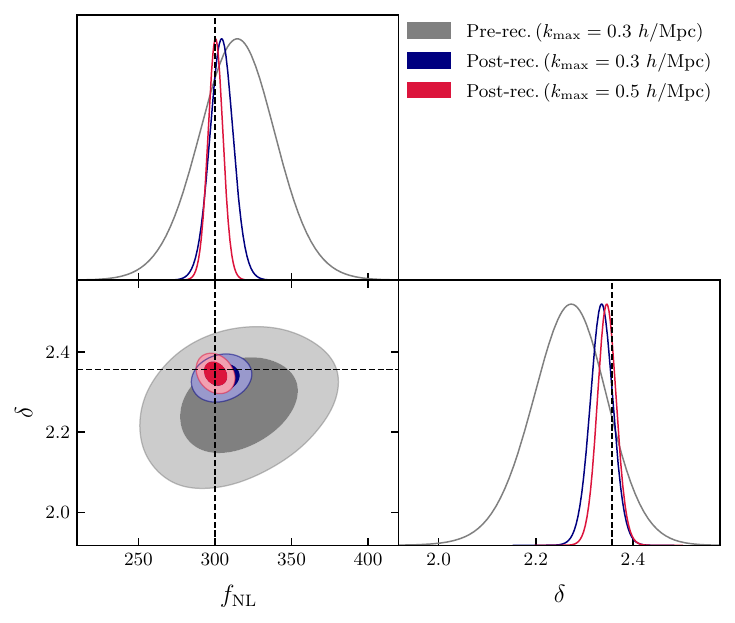}

\caption{Comparison of constraints on $f_{\rm NL}$ and $\delta$ using the pre- and post-reconstruction squeezed matter bispectrum. Assuming $k_{\rm max}=0.3~h/{\rm Mpc}$, reconstruction leads to a factor of 3.3 improvement in the constraint on $f_{\rm NL}$ and $\delta$. Moreover, with the post-reconstruction bispectrum, we can extend the maximum wavenumber to $k_{\rm max}=0.5~h/{\rm Mpc}$ for which we find a factor of 5.2 (4.0) improvement in the constraint on $f_{\rm NL}$ ($\delta$) compared to the pre-reconstruction results. We do not show the constraints from the pre-reconstruction bispectrum with $k_{\rm max}=0.5~h/{\rm Mpc}$ because these lead to negligible improvements over the $k_{\rm max}=0.3~h/{\rm Mpc}$ results.}
\label{fig:pre_post_recon_constraints}
\end{figure}

Finally, Fig.~\ref{fig:pre_post_recon_constraints} compares the constraint on $f_{\rm NL}$ and $\delta$ using the pre- and post-reconstruction bispectrum measurements at $z=0$. We fit the pre-reconstruction bispectrum up to $k_{\rm max}=0.3~h/{\rm Mpc}$ since the oscillations are damped beyond these scales and we find negligible improvement increasing $k_{\rm max}$ further. On the other hand, we present constraints from the post-reconstruction bispectrum assuming $k_{\rm max}=0.3~{h/{\rm Mpc}}$ (blue) and $k_{\rm max}=0.5~{h/{\rm Mpc}}$ (red). Not only do we find that we are able to recover unbiased constraints on $f_{\rm NL}$ and $\delta$ using the $z=0$ post-reconstruction squeezed matter bispectrum, but we also find significant improvement in our constraints. For example, the error on $f_{\rm NL}$ shrinks from 26 to $8$ (5) assuming $k_{\rm max}=0.3~h/{\rm Mpc}$ ($0.5~h/{\rm Mpc}$). Thus, we find that reconstruction leads to  $5\times$ improvement in constraints on oscillatory features in the squeezed bispectrum.

For comparison, we perform a similar analysis on the pre- and post-reconstruction bispectra from simulations with local primordial non-Gaussianity using the Quijote simulations with $f_{\rm NL}^{\rm loc}=100$ with the same bins and scale cuts. We again find unbiased constraints on $f_{\rm NL}^{\rm loc}$ from the pre- and post-reconstrution bispectrum, reinforcing our conclusion that the post-reconstruction squeezed matter bispectrum can be accurately modeled using reconstructed separate universe simulations. On the other hand, we find that the constraint on $f_{\rm NL}^{\rm loc}$ from the post-reconstruction bispectrum with $k_{\rm max}=0.3~h/{\rm Mpc}$ ($0.5~h/{\rm Mpc}$) is only a factor 2.0 (2.6) tighter than the constraint from the pre-reconstruction bispectrum with $k_{\rm max}=0.3~h/{\rm Mpc}.$ These results are quantitatively similar with previous studies of reconstruction in the context of $f_{\rm NL}^{\rm loc}$~\cite{Shirasaki:2020vkk, Floss:2023ylq, Chen:2024exy, Bottema:2025vww}. 

The reason that reconstruction is significantly more effective for oscillatory signals is because reconstruction has two effects on the oscillatory sims. First, it enhances the primordial signal by undoing much of the damping of the oscillations (see Fig.~\ref{fig:full_bispectrum_meas_Delta0}). Importantly, this suggests that the damping of oscillatory features in the squeezed bispectrum is largely due to large-scale bulk flows, as opposed to small scale non-linearities. This is consistent with the agreement between Zel'dovich and the response functions (Fig.~\ref{fig:response_function}). Second, reconstruction reduces the significant off-diagonal correlations between hard-mode bins. For local $f_{\rm NL}$, only the second effect is relevant, leading to the reduced improvements in $\sigma(f_{\rm NL})$ sourced by reconstruction.

Whereas the fact that the damping of oscillations in the squeezed bispectrum is predominately sourced by long-wavelength displacements is encouraging for searching for oscillatory features in the post-reconstruction bispectrum, it could have important implications for the Cosmological Collider scenario. In particular, the separate universe transformation for the Cosmological Collider scenario includes an additional $k^{-3/2}$ rescaling of the power spectrum, which significantly alters the long-wavelength power spectrum, and, in turn, the non-linear evolution of the oscillations. While a detailed study of this effect is beyond the scope of this work, in Appendix~\ref{App:Delta_gtr_0} we compute separate universe response functions with the $k^{-3/2}$ scaling and find that they are \emph{extremely} sensitive to changes in the long-wavelength linear power spectrum used in the simulations. This could present a challenge for robustly modeling the non-linear evolution of oscillations in the Cosmological Collider scenario, indicating the limitations of the separate universe approach.

\section{Conclusion}\label{Sec:conclusions}

\noindent Oscillations in the primordial bispectrum arise in a range of well-motivated early Universe scenarios. LSS surveys provide a promising avenue to probe these features. However, extracting such signals from LSS data requires a detailed understanding of how non-linearities impact such oscillations. In this work, we have taken an important first step towards understanding the non-linear evolution of oscillations in the bispectrum by running a suite of $N$-body simulations with an oscillatory primordial bispectrum. Using these simulations, we analyzed the non-linear evolution of oscillations in the squeezed matter bispectrum. Our main findings are as follows:
\begin{itemize}
    \item  Non-linear evolution significantly damps oscillations in the squeezed bispectrum. For the model studied herein (Eq.~\ref{eq:squeezed_bk_osc}), the oscillatory features are almost entirely erased by $k\approx 0.3~h/{\rm Mpc}$ at $z=0$, though the damping is less severe at higher redshifts.
    \item In the quasi-linear regime, the damping in the squeezed bispectrum is well-described by the Zel'dovich approximation, suggesting that it is predominately sourced by long-wavelength displacements. In the fully non-linear regime, separate universe simulations with oscillatory features in the primordial power spectrum provide an accurate description of the damping.
    \item By combining the bispectrum measurements with numerical derivatives extracted from separate universe simulations, we derive unbiased constraints on the amplitude and phase of primordial oscillatory features from measurements of the squeezed matter bispectrum deep into the non-linear regime.
    \item The iterative reconstruction algorithm from \citet{Schmittfull:2017uhh} can largely restore oscillatory features in the squeezed matter bispectrum. Moreover, the post-reconstruction squeezed bispectrum can be accurately modeled by applying the same reconstruction algorithm to the separate universe simulations, yielding up to a \emph{fivefold} improvement in constraints on oscillatory features in the bispectrum compared to the pre-reconstruction case, at $z=0$. This gain is roughly twice as large as that achieved for simulations with local non-Gaussianity, highlighting the power of reconstruction at simultaneously enhancing the signal and reducing the non-Gaussian covariance for oscillatory bispectra.
\end{itemize}

There are several promising avenues for further exploration. A natural next step would be to run $N$-body simulations including the additional $(q/k)^{-3/2}$ suppression characteristic of the Cosmological Collider scenario. Compensating for this suppression would require larger simulation volumes and/or more realizations than used in the present work. Moreover, this steep scaling may introduce new challenges for modeling the non-linear evolution of this signal using the separate universe approach. In particular, the separate universe transformations for this scenario significantly alter the long-wavelength power spectrum, and hence the non-linear evolution of the oscillations in the response functions (see Appendix~\ref{App:Delta_gtr_0} for more details). While a detailed investigation of this scenario is beyond the scope of this work, these considerations suggest that accurately predicting the non-linear evolution of oscillatory signatures in Cosmological Collider models may require a more nuanced treatment than what is done here, \emph{e.g.}, by studying the bispectrum beyond the squeezed limit.

Similarly, it would be valuable to study the non-linear evolution and reconstruction of oscillatory features in non-squeezed bispectrum configurations, particularly those motivated by primordial features models, which induce linear or logarithmic oscillations in the equilateral bispectrum. 

On the observational side, our results could be used to develop new estimators to constrain oscillatory bispectra from weak lensing and galaxy clustering data deep into the non-linear regime~\cite{Giri:2023mpg, Anbajagane:2023wif, Goldstein:2023brb}. These methods would be highly complementary to perturbative approaches, which have recently been used to constrain primordial non-Gaussianity from galaxy surveys~\cite{DAmico:2022gki, Cabass:2022ymb, Cabass:2022wjy, Ivanov:2024hgq, Cabass:2024wob, Chaussidon:2024qni}. A key step in this direction would be to understand how oscillatory primordial bispectra impact galaxy statistics, such as the scale-dependent bias and the galaxy bispectrum~\cite{Cyr-Racine:2011bjz, MoradinezhadDizgah:2017szk, MoradinezhadDizgah:2018ssw, Cabass:2018roz}. To this end, we note that our separate universe simulations can be directly used to estimate the non-Gaussian bias parameters associated with the oscillatory bispectrum in Eq.~\eqref{eq:squeezed_bk_osc}~\cite{Cabass:2018roz, Goldstein:2024bky}. Moreover, although this paper focused on the squeezed matter bispectrum, it is straightforward to generalize our methodology to study oscillatory features in the collapsed matter trispectrum~\cite{Giri:2023mpg, Goldstein:2024bky}.

To apply these results to realistic spectroscopic galaxy survey data, we would need to account for galaxy bias and redshift-space distortions (RSD). While both effects may significantly complicate the modeling of the squeezed bispectrum, they do not violate the LSS consistency conditions~\cite{Creminelli:2013poa, Nishimichi:2015kla}. Consequently, any observation of a pole in the squeezed redshift-space galaxy bispectrum would be a smoking gun signature of new physics. However, translating the presence (absence) of such a pole into a detection (constraint) on $f_{\rm NL}$ would require accounting for galaxy bias and RSD. To account for galaxy bias, one would need to estimate the response of the \emph{galaxy} power spectrum to a long-wavelength potential perturbation~\cite{Giri:2023mpg}, which could be highly dependent on the galaxy sample, analogous to the well-known results for scale-dependent bias~\cite{Barreira:2020kvh, Barreira:2022sey, Lazeyras:2022koc}. Though work would also be needed to model non-linear RSD, a possible workaround would be to construct galaxy samples with reduced non-linear contributions~\cite{BaleatoLizancos:2025wdg}.

Given the promising results from the post-reconstruction bispectrum analysis, it would be worthwhile to systematically investigate how different assumptions about the reconstruction algorithm and post-reconstruction modeling affect the recovery of oscillatory features in the squeezed bispectrum. In this regard, recent developments in Lagrangian perturbation theory modeling for the bispectrum could be particularly valuable~\cite{Chen:2024pyp}. It would also be important to investigate how well reconstruction works for oscillations in the galaxy bispectrum, where additional complications arise from galaxy bias, RSD, and shot noise. Finally, it could be interesting to study oscillatory bispectra using field-level inference, which has been shown to be a powerful probe of oscillatory features in the power spectrum~\cite{Babic:2022dws, Babic:2024wph}.

Ultimately, this work demonstrates that, while they may wriggle their way out of standard analyses, non-perturbative treatments will leave little wiggle room for early Universe wiggles.

\acknowledgments

\noindent We thank Dhayaa Anbajagane, Colin Hill, and Beatriz Tucci for useful discussions.  SG acknowledges support from NSF grant AST-2307727. OHEP is a Junior Fellow of the Simons Society of Fellows, and thanks Jane Street for their Uber credits. The authors acknowledge the Texas Advanced Computing Center (TACC)\footnote{\href{http://www.tacc.utexas.edu}{http://www.tacc.utexas.edu}} at The University of Texas at Austin for providing computational resources that have contributed to the research results reported within this paper. We acknowledge computing resources from Columbia University's Shared Research Computing Facility project, which is supported by NIH Research Facility Improvement Grant 1G20RR030893-01, and associated funds from the New York State Empire State Development, Division of Science Technology and Innovation (NYSTAR) Contract C090171, both awarded April 15, 2010.

\bibliographystyle{apsrev4-1}
\bibliography{biblio}

\onecolumngrid
\clearpage
\appendix

\begin{figure*}[!t]
\centering
\includegraphics[width=0.995\linewidth]{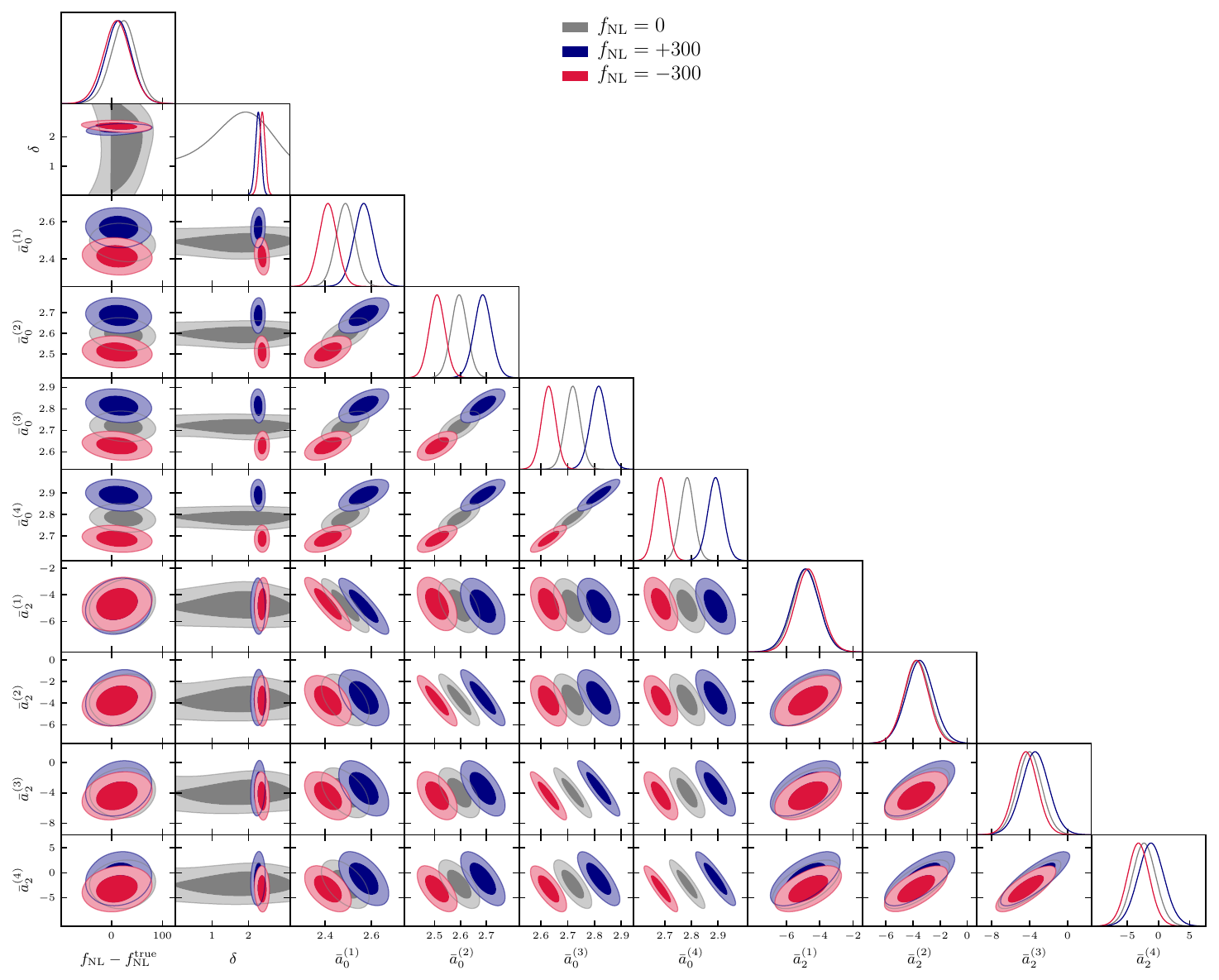}
\caption{Marginalized posterior distribution for all primordial and gravitational non-Gaussianity parameters from a joint analysis of the squeezed matter bispectrum and long wavelength matter power spectrum at $z=0,$ assuming $k_{\rm max}=0.3~h/{\rm Mpc}$. Notice that $f_{\rm NL}$ values of $\pm 300$ introduce an $\approx \pm 3\%$ shift in $\bar{a}_0^{(i)}$ due to the coupling of primordial non-Gaussianity and gravity. We account for this shift when producing Fig.~\ref{fig:full_bispectrum_meas_Delta0} using a procedure outlined in Appendix~\ref{App:O_fnlsq_contribution}.}
\label{fig:full_posterior}
\end{figure*}
\section{Higher-order contributions to the power spectrum and bispectrum}\label{App:O_fnlsq_contribution}
\subsection{Contributions from the coupling of primordial non-Gaussianity to gravity}

\noindent To qualitatively assess the damping of oscillations in the squeezed matter bispectrum we computed the following statistic (Eq.~\ref{eq:Bspec_differnce}),
\begin{equation}\label{eq:Bspec_differnce_copied}
    \frac{\Delta B(q,k)}{P(q)P(k)}\equiv \frac{B^{f_{\rm NL}^\pm}_m(q,k)}{P^{f_{\rm NL}^\pm}_m(q)P^{f_{\rm NL}^\pm}_m(k)}-\frac{B^{G}_m(q,k)}{P^G_m(q)P^{G}_m(k)}.
\end{equation}
Although this fully cancels the purely gravitational contributions to the bispectrum, it includes contributions from the coupling of primordial non-Gaussianity to gravity, which are \emph{a priori} difficult to model. Specifically, using Eq.~\eqref{eq:B_squeezed_full}, non-perturbative model for Eq.~\eqref{eq:Bspec_differnce_copied} is given by
\begin{equation}\label{eq:delta_B_squeezed_full}
\begin{split}
    \frac{\Delta B(q,k)}{P(q)P(k)}=&\frac{3f_{\rm NL}\Omega_mH_0^2}{D_{\rm md}(z)\,q^{2}T(q)}\bigg[\cos(\mu\ln q+\delta) \frac{\partial\ln P_m(k|\mu)}{\partial \epsilon_c}\bigg\vert_{\epsilon_c=0}+\sin(\mu\ln q+\delta) \frac{\partial \ln P_m(k|\mu)}{\partial \epsilon_s}\bigg\vert_{\epsilon_s=0}\bigg] \\
&+\left(\Delta a_0(k)+\Delta {a}_2(k)\frac{q^2}{k^2}+\cdots\right)+\mathcal{O}(f_{\rm NL}^2),
\end{split}
\end{equation}
where $\Delta a_0(k)$ and $\Delta a_2(k)$ represent the $\mathcal{O}(f_{\rm NL})$ corrections to $a_0(k)$ and $a_2(k)$ that arise from the coupling of primordial non-Gaussianity to gravity.

To quantify how large these corrections are, Fig.~\ref{fig:full_posterior} shows marginalized posterior for $f_{\rm NL}$, as well as all gravitational non-Gaussianity nuisance parameters from a joint analysis of the squeezed matter bispectrum and long wavelength matter power spectrum at $z=0$, assuming $k_{\rm max}=0.3~{h/{\rm Mpc}}.$ By comparing the constraints on $\bar{a}_0^{(i)}$ for simulations with and without primordial non-Gaussianity, we observe that our simulations with $f_{\rm NL}=\pm 300$ lead to an $\approx\pm 3\%$ shift in $\bar{a}_0^{(i)}$.  In contrast, for the simulation volume and scale cuts used here, we do not observe a statistically significant shift in the $\bar{a}_2^{(i)}$ posteriors, which are constrained with considerably less precision than $\bar{a}_0^{(i)}$.

Since the gravitational contribution to the squeezed bispectrum at $z=0$ dominates over the primordial signal in our simulations, the small shifts in $\bar{a}_0^{(i)}$ from non-zero values of $f_{\rm NL}$ indicate that the bispectrum difference measurements (Eq.~\ref{eq:Bspec_differnce_copied}) cannot be directly compared to theoretical predictions using only the leading term in Eq.~\eqref{eq:delta_B_squeezed_full}. In principle, one could fit for these nuisance parameters directly, which is exactly what we do in our coarse-binned analysis. However, given the large data-vector used in Fig.~\ref{fig:full_bispectrum_meas_Delta0} and the significant correlations between bispectrum measurements as a function of $k$, this would require \emph{significantly} more simulations to estimate the covariance\footnote{Note that we cannot simply use more realizations of the \textsc{Quijote} simulations with Gaussian initial conditions to estimate the covariance required for this problem because we are directly estimating the difference in the bispectrum between simulations with and without non-Gaussianity.}.

Instead, in Fig.~\ref{fig:full_bispectrum_meas_Delta0}, we apply a small simulation-based correction for visualization purposes. Specifically, since the $\mathcal{O}(f_{\rm NL})$ contributions to $\bar{a}_{2}^{i}$ are negligible for the simulation volume and scale cuts analyzed here (see Fig.~\ref{fig:full_posterior}), the contributions from the coupling $f_{\rm NL}$ to gravity to $\Delta B(q,k)/[{P(q)P(k)}]$ should be approximately independent of $q$. Consequently, we estimate this offset by taking the difference between our theoretical and measured bispectrum, assuming the fiducial values of $f_{\rm NL}$ and $\delta$, but $\bar{a}_0^{(i)}=\bar{a}_2^{(i)}=0$, and computing the mean averaged over the four $q$-bins in Fig.~\ref{fig:full_bispectrum_meas_Delta0}. We subtract this average from the $f{\rm NL}=+300$ measurements and add the same average to the $f_{\rm NL}=-300$ measurements. After accounting for this correction,\footnote{This correction is negligible at $z=3$, but becomes important at $z=0$, particularly at high-$k$ where the primordial oscillations are exponentially suppressed.} we find excellent agreement between the theory model and measured bispectra across all redshifts and scales. Notably, the consistency of this correction across $q$-bins, and the fact that the correction derived from $f_{\rm NL}=+300$ simulations accurately describes the $f_{\rm NL}=-300$ simulations, provides a strong, non-trivial validation of our estimate of the contribution arising from the coupling of primordial non-Gaussianity to gravity.

\subsection{$\mathcal{O}(f_{\mathrm{NL}}^2)$ contributions to the power spectrum and bispectrum}
\noindent To quantify the size of $\mathcal{O}(f_{\rm NL}^2)$ corrections to the bispectrum, we compute the difference between the output of the simulations with positive and negative $f_{\rm NL}$,
\begin{equation}\label{eq:Bspec_difference_symm}
    \frac{\Delta B(q,k)}{P(q)P(k)}_{\rm sym}\equiv \frac{1}{2}\times\left(\frac{B^{f_{\rm NL}^+}_m(q,k)}{P^{f_{\rm NL}^+}_m(q)P^{f_{\rm NL}^+}_m(k)}-\frac{B^{f_{\rm NL}^-}_m(q,k)}{P^{f_{\rm NL}^-}_m(q)P^{f_{\rm NL}^-}_m(k)}\right),
\end{equation}
and compare it to the difference defined in Eq.~\eqref{eq:Bspec_differnce_copied}. Since the two expressions agree at $\mathcal{O}(f_{\rm NL})$, any difference indicates higher-order terms in the analysis.

\begin{figure*}[!t]
\centering
\includegraphics[width=0.995\linewidth]{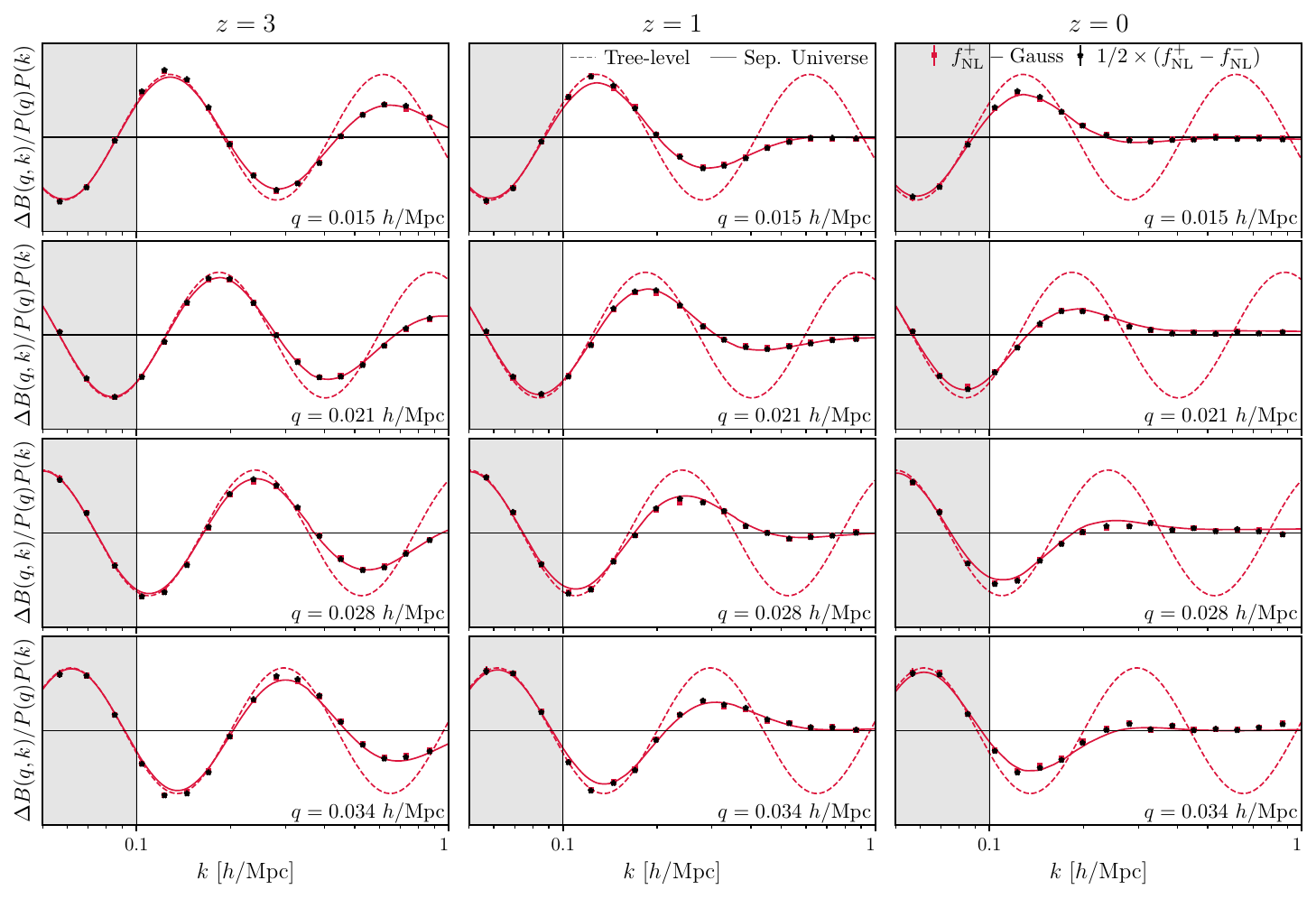}
\caption{Impact of $\mathcal{O}(f_{\rm NL}^2)$ contributions on the bispectrum-power spectrum ratio, the relevant quantity for the analyses presented in this work. The red points are measured by taking the difference between simulations with $f_{\rm NL}=+300$ and Gaussian initial conditions (Eq.~\ref{eq:Bspec_differnce_copied}). The black points are measured by taking the difference between simulations with $f_{\rm NL}=+300$ and $f_{\rm NL}=-300$ (Eq.~\ref{eq:Bspec_difference_symm}). The two procedures give consistent results within the errorbars, thus $\mathcal{O}(f_{\rm NL}^2)$ corrections are negligible for this work.}
\label{fig:bispectrum_full_Ofnl_sq_corrections}
\end{figure*}

Fig.~\ref{fig:bispectrum_full_Ofnl_sq_corrections} shows a comparison of bispectrum-power spectrum ratio as estimated by both methods. The two approaches give consistent results within the statistical errors, thus this particular statistic is robust to $\mathcal{O}(f_{\rm NL}^2)$ corrections for the simulation settings used in this work. Note, however, that this consistency check does not necessarily imply that $\mathcal{O}(f_{\rm NL}^2)$ contributions to the bispectrum are negligible, as these contributions partially cancel when taking the ratio. Nevertheless, this ratio is the relevant quantity to assess in the present work, since we form a joint likelihood using the bispectrum and power spectrum measured from a particular simulation, and compute the response functions by weighting the power spectrum of this simulation with the logarithmic derivative estimated from the separate universe simulations. In other words, the likelihood is defined consistently in terms of $B(q,k)$, $P(q)$, and $P(k)$ estimated from a particular realization so that it is effectively only sensitive to the ratio in Eq.~\eqref{eq:Bspec_differnce_copied}. See Sec.~\ref{subsec:measurements_modelling_likelihood} for more details.

\vfill
\clearpage
\pagebreak

\begin{figure*}[!t]
\centering
\includegraphics[width=0.995\linewidth]{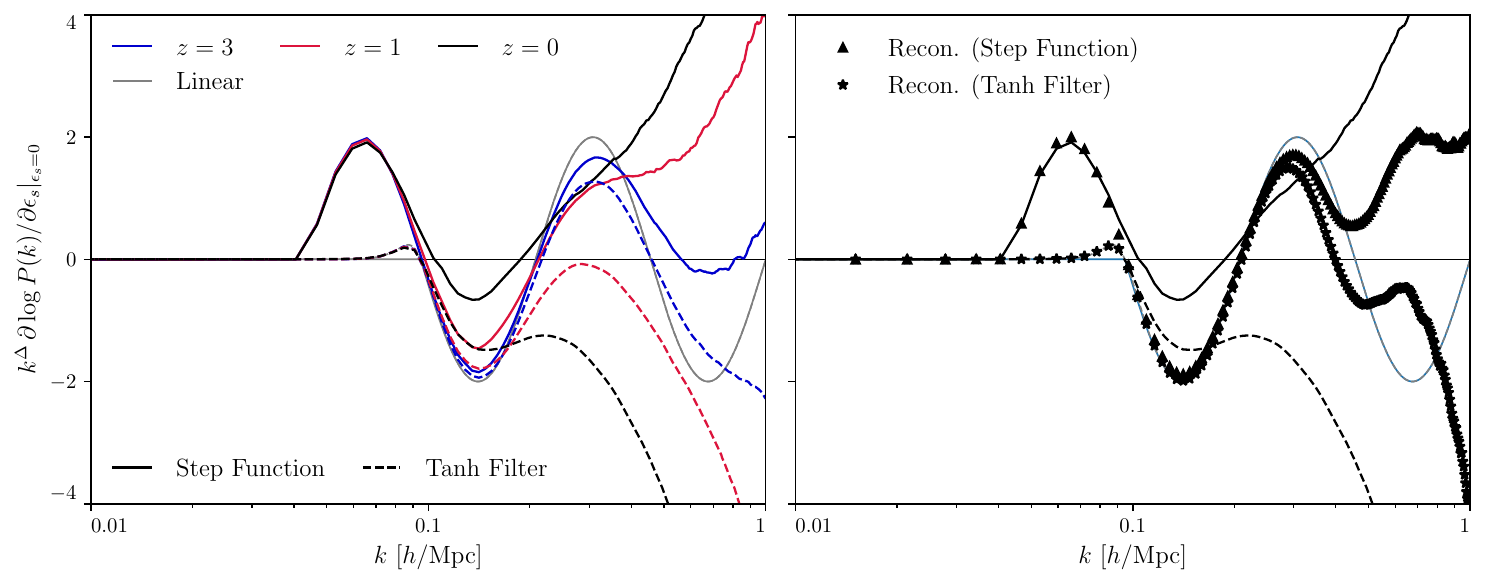}
\caption{Non-linear contribution to the potential derivative response function evaluated using separate universe simulations with $\Delta=3/2$. The left panel shows the redshift evolution of the potential derivative. As described in the text, the solid (dashed) lines correspond to separate universe simulations with a step function (tanh -- Eq.~\ref{eq:tanh_filter}) high-pass filter to regulate the modifications to the large-scale power spectrum. Despite the relatively small difference in filters, the non-linear response functions differ significantly between the two cases. The right panel shows the $z=0$ response functions before (lines) and after (markers) reconstruction. Whereas the pre-reconstruction response functions are extremely sensitive to the choice of filter, the post-reconstruction response functions are remarkably robust. Moreover, the reconstructed response functions have prominent oscillatory features, suggesting that reconstruction could be highly effective in the $\Delta=3/2$, as was found with $\Delta=0$ analysis in the main text. Note that these response functions are noisier than those used in the main text (Fig.~\ref{fig:response_function}), as they are estimated using only two realizations.}
\label{fig:SU_response_Delta_1p5}
\end{figure*}
\section{Non-perturbative evolution of oscillatory bispectra with $\Delta=3/2$}\label{App:Delta_gtr_0}
\noindent In this appendix, we discuss several challenges associated with analyzing the non-perturbative evolution of the squeezed bispectrum in Eq.~\eqref{eq:b_sq_generic} for the Cosmological Collider Scenario where $\Delta=3/2.$ Whereas we could use the methods in this work to runs simulations with $\Delta=3/2$, this is computationally challenging since we would need to run simulations with larger volumes and/or more realizations compensate for the $(q/k)^{3/2}$ suppression. Note that simply increasing the value of $f_{\rm NL}$ is insufficient because this would lead to large $\mathcal{O}(f_{\rm NL}^2)$ contributions. Given these challenges, we defer a detailed study of the $\Delta=3/2$ scenario here to future work. Nevertheless, we can garner some insight into the non-perturbative evolution of the Cosmological Collider bispectrum using separate universe simulations. This is the focus of the present appendix.

For $\Delta=3/2$, the separate universe transformations (Eqs.~\ref{eq:SU_transformation_cos} and~\ref{eq:SU_transformation_sin}) rescale the power spectrum by $k^{-3/2}$, leading to significant modifications to the power spectrum at both long and short wavelengths. This introduces several complications. First, there is a computational challenge: for $\Delta = 3/2$, the transformation varies by roughly three orders of magnitude across the range of scales we study ($0.01 \lesssim k \lesssim 1~h/{\rm Mpc}$). At the resolution and volume of our simulations, it is not possible to choose $\epsilon_c$ and $\epsilon_s$ such that the transformation remains a small but resolvable perturbation across all scales. One possibility to circumvent this is to only modify the power spectra of the separate universe simulations above some scale $k>k_*$ by introducing a high-pass filter. Formally, this procedure is 
well-motivated as the separate universe procedure assumes we are working in a patch that is much smaller than the background mode, $q$. However, since the non-linear evolution of oscillatory features is dominated by long-wavelength displacements, the response functions will depend heavily on the shape and scale of this filter. 

To illustrate this, we run two sets of separate universe simulations for the sine response function (Eq.~\ref{eq:SU_transformation_sin}) and $\Delta=3/2$. In the first set of simulations, we apply a sharp step function that modifies modes only above $k_*^{\rm step} \approx 0.04~h/{\rm Mpc}$, where $k_*$ is set to zero of the separate universe transformation closest to $k = 0.05~h/{\rm Mpc}$, ensuring continuity. In the second set, we apply a smooth filter,
\begin{equation}\label{eq:tanh_filter}
    F(k)=\frac{1}{2}\left[1+\tanh\left( \frac{k-k_*}{\Delta k}-1\right) \right],
\end{equation}
where $k_*=0.08~h/{\rm Mpc}$ and $\Delta k =0.01~h/{\rm Mpc}.$ In both cases, we run two realizations each and fix $\epsilon_s = 0.003~[{\rm Mpc}/h]^{3/2}$.\footnote{We ran simulations with several values of $\epsilon$ to ensure convergence.}

Fig.~\ref{fig:SU_response_Delta_1p5} compares the resulting response functions, with the redshift evolution shown in the left panel. Notably, the non-linear evolution is highly sensitive to the filter choice. At $z=0$, the sign of the response differs above $k \gtrsim 0.2~h/{\rm Mpc}$. Since the choice of cutoff depends on the long wavelength mode $q$ and on the transition of the bispectrum from squeezed to non-squeezed configurations, these findings could suggest that a more careful understanding of this transition, which is generally model dependent, may be necessary to accurately model the squeezed non-linear evolution of oscillatory features in the Cosmological Collider scenario. More broadly, these results might simply reflect the limited applicability of separate universe simulations in models where the squeezed limit of the primordial bispectrum does not differ substantially from that of the gravitational bispectrum.

Nevertheless, all hope of searching for these oscillatory features in the non-linear squeezed bispectrum may not be lost. Since the damping is predominately sourced by long-wavelength effects, it seems likely that the post-reconstruction bispectrum is less sensitive to the treatment of long-wavelength modes. As a preliminary test of this, the right panel shows the $z=0$ response functions before and after reconstruction. While the pre-reconstruction responses differ significantly, the post-reconstruction results agree well with each other (and linear theory) up to $k \approx~0.3~h/{\rm Mpc}$. It is encouraging that reconstruction effectively recovers the oscillatory signal, despite the steep $k^{3/2}$ scaling.

Ultimately, this appendix demonstrates that more work is necessary to understand the non-linear evolution of oscillations when $\Delta=3/2$. 
\end{document}